\pdfoutput=1
\documentclass[11pt]{article}

\usepackage[utf8]{inputenc}
\usepackage[T1]{fontenc}
\usepackage[english]{babel}
\usepackage{lmodern}
\usepackage[letterpaper,margin=1in]{geometry}

\usepackage{amsmath}
\usepackage{amssymb}
\usepackage{amsfonts}
\usepackage{graphicx}
\usepackage{xcolor}
\usepackage{booktabs}
\usepackage{placeins}
\usepackage{authblk}
\usepackage[font=small,labelfont=bf]{caption}
\usepackage[numbers,sort&compress]{natbib}
\usepackage{microtype}
\usepackage{orcidlink}

\usepackage{hyperref}
\usepackage{url}

\hypersetup{%
    colorlinks=true,
    allcolors=blue,
    breaklinks=true,
    pdfauthor={Michael Acquah and Zheng Liu},
    pdftitle={Generative Design of Liquid-Cooling Channels for Thermal Management of 2.5D and 3D Integrated Advanced Packaging},
    pdfkeywords={Generative design, advanced electronic packaging, liquid cooling, thermal management, conjugate heat transfer},
    pdfsubject={arXiv preprint},
}

\newlength{\nomlabelwidth}
\newenvironment{nomenclature}%
  {\section*{Nomenclature}\begingroup\raggedright\small}%
  {\endgroup}
\newcommand{\EntryHeading}[1]{\par\medskip\noindent\textbf{#1}\par\smallskip}
\newcommand{\entry}[2]{%
  \par\noindent
  \makebox[\nomlabelwidth][l]{#1}%
  \parbox[t]{\dimexpr\linewidth-\nomlabelwidth\relax}{#2}\par\vspace{1pt}}

\providecommand{\keywords}[1]{\par\vspace{0.6em}\noindent\textbf{Keywords:} #1}

\title{\bfseries Generative Design of Liquid-Cooling Channels for
Thermal Management of 2.5D and 3D Integrated Advanced Packaging}

\author[1]{Michael Acquah \orcidlink{0009-0006-7944-860X}}
\author[2]{Zheng Liu\thanks{Corresponding author: \href{mailto:zhengtl@umich.edu}{zhengtl@umich.edu}} \orcidlink{0000-0003-4869-8893}}
\affil[1]{Department of Computer and Information Science, University of Michigan-Dearborn,
4901 Evergreen Rd, Dearborn, MI 48128, USA}
\affil[2]{Department of Industrial and Manufacturing Systems Engineering,
University of Michigan-Dearborn, Dearborn, MI 48128, USA}

\date{}

\begin{document}

\maketitle

\begin{abstract}
\noindent
High-power multi-chip packages require increasingly effective cooling as intensive heat is generated within a limited package area. This work presents a physics-guided generative design framework for liquid-cooling channel topology optimization in a $2.7~\mathrm{kW}$ multi-chip package containing two high-power graphics processing units (GPUs) and one central processing unit (CPU). A conditional diffusion model generates symmetric channel layouts using maximum GPU temperature, GPU temperature spread, and pressure drop as performance targets. Generated designs undergo connectivity and dead-end-branch screening and are evaluated using a calibrated reduced-order thermal-fluids model. Based on 5,000 generated layouts, the multi-objective analysis identified the optimal design for thermal properties, with estimated maximum GPU temperature of $70.30^\circ\mathrm{C}$, GPU temperature spread of $24.90^\circ\mathrm{C}$, and pressure drop of $89.72~\mathrm{kPa}$. Compared to a conventional reference topology, the optimal design reduced the maximum GPU temperature, temperature spread, and pressure drop by $33.6\%$, $52.5\%$, and $72.8\%$, respectively. High-fidelity three-dimensional conjugate heat-transfer simulation in OpenFOAM estimated a maximum GPU temperature of $66.70^\circ\mathrm{C}$ and a pressure drop of $92.1~\mathrm{kPa}$, showing only differences of $8.6\%$ in temperature rise and $2.6\%$ in pressure drop. The results demonstrate that physics-guided generative design based on the reduced-order model can efficiently discover unconventional cooling channel architectures while reducing reliance on repeating computationally expensive simulation.

\keywords{Generative design, advanced electronic packaging, liquid cooling, thermal management, conjugate heat transfer}
\end{abstract}

\section{Introduction}

The continued increase in power density of electronic packages has made thermal management a critical constraint in the design of advanced computing systems. As processors, graphics processing units, and heterogeneous chiplets are integrated within compact multi-chip packages, heat must be removed over smaller areas while maintaining acceptable junction temperatures, temperature uniformity, and system reliability. Conventional air-cooling solutions become increasingly limited under high heat-flux conditions due to their relatively low heat capacity and limited ability to remove heat from densely packed components. Liquid cooling offers a more compact and effective alternative because the coolant can be routed close to the heat-generating devices, increasing convective heat removal while reducing the thermal resistance between the chip and the heat sink \cite{gullbrand2019liquid}.

Microchannel and cold-plate cooling have therefore received significant attention for high-power electronic cooling. The foundational work of Tuckerman and Pease demonstrated the potential of integrated water-cooled microchannel heat sinks for very high heat-flux removal in silicon devices~\cite{tuckerman1981high}. Subsequent studies investigated three-dimensional (3D) flow and heat transfer in microchannel heat sinks, showing that channel geometry, wall conduction, flow rate, and pressure drop strongly influence the resulting thermal performance~\cite{qu2002analysis,qu2002experimental,zhao2002analysis, wei2004stacked}. Reviews of single-phase microchannel cooling further established that compact liquid-cooled heat sinks can provide high heat-transfer coefficients, but often at the cost of increased hydraulic resistance and pump power~\cite{kandlikar2003evolution,garimella2004single,morini2004single,asadi2014review}. This trade-off between thermal performance and pressure loss remains central to the design of liquid-cooling channels for electronic packages.

For multi-chip electronic packages, the cooling problem is more complex than removing heat from a single uniform source. Multiple heat-generating components may have different footprints, power levels, and spatial locations, leading to nonuniform heat distribution and localized hot spots. A simple straight-channel cold plate is easy to manufacture and may offer a low pressure drop, but it may not distribute coolant effectively across all chip regions. Flow maldistribution within parallel microchannel networks can further increase temperature nonuniformity, motivating geometric modifications that improve coolant distribution while introducing additional hydraulic trade-offs~\cite{kumar2019numerical}. Several quantification methods based on velocity, mass-flow, and temperature nonuniformity have been proposed to standardize comparisons of maldistribution severity across cooling-channel designs~\cite{dkabrowski2023comparison}. More advanced layouts, including branching, manifold, and multi-pass channel networks, can improve coolant distribution and local thermal coverage. Modifications to conventional straight-channel geometries have also been investigated to improve thermal-fluids performance; for example, Xie et al. demonstrated that wavy microchannel configurations can alter both heat-transfer performance and pressure-drop characteristics compared with straight channels \cite{xie2012analysis}. Recent embedded 3D manifold microchannel architectures have further demonstrated the potential of distributed coolant delivery for high-power electronic devices, with thermal performance and pressure loss strongly dependent on manifold and channel layout~\cite{lin2024multi}. Dede demonstrated that multipass-branching microchannel heat sinks can be designed for high-heat-flux electronics cooling using optimization-based methods~\cite{dede2012optimization}. Related experimental and numerical work on branching microchannel heat sinks showed that such designs can provide favorable thermal-fluid performance when properly configured~\cite{dede2013experimental}. However, manually designing these layouts is difficult because increasing channel complexity expands the design space and introduces competing objectives, including maximum and average temperature, pressure drop, pumping power, and flow uniformity.

Topology optimization has been widely explored as a systematic approach for designing fluid and thermal-fluid systems. Early work by Borrvall and Petersson formulated topology optimization for Stokes flow~\cite{borrvall2003topology}, while Gersborg-Hansen et al. extended such methods to channel-flow problems~\cite{gersborg2005topology}. Later studies incorporated more complex thermal-fluid effects, including coupled heat transfer and turbulent flow~\cite{yaji2015topology, dilgen2018topology}. In electronic cooling, topology optimization has been applied to generate non-intuitive cold-plate and heat-sink geometries that improve heat removal or temperature uniformity under hydraulic constraints. These approaches provide a rigorous optimization framework, but they can be computationally expensive and may require repeated solution of governing flow and energy equations. In addition, designs produced by continuous topology optimization often require post-processing to convert them into connected, manufacturable channel geometries.

Recent advances in machine learning and generative modeling provide an opportunity to accelerate the design of cooling-channel layouts. Instead of evaluating every candidate design with high-fidelity computational fluid dynamics (CFD), surrogate models can be trained to approximate thermal or flow responses at a much lower computational cost. Convolutional neural networks have been used to approximate steady flow fields and other physics-governed spatial fields~\cite{guo2016convolutional}. The U-Net architecture, originally developed for image-to-image prediction tasks, is particularly well-suited to field prediction because its encoder-decoder structure combines global context with spatial localization~\cite{ronneberger2015u}. For cooling-channel design, a U-Net surrogate can map a channel-mask representation to a estimated temperature field, enabling rapid screening of many candidate geometries. Physics-informed and physics-guided learning approaches further motivate the integration of physical constraints with neural network-based prediction~\cite{raissi2019physics}. Generative adversarial networks have also been combined with CNN-based surrogates to optimize heat sink geometries directly, though such approaches can suffer from training instability and mode collapse relative to diffusion-based alternatives~\cite{flynn2023heat}.

At the same time, diffusion-based generative models have become powerful tools for generating complex spatial patterns. Denoising diffusion probabilistic models generate samples by learning to reverse a gradual noising process~\cite{ho2020denoising}, and score-based generative modeling provides a related continuous-time formulation using stochastic differential equations~\cite{song2020score}. Although these models were initially developed for image generation, their ability to learn distributions over structured fields makes them useful for engineering design problems represented as images or masks. Conditional diffusion models have been shown to outperform generative adversarial networks for topology optimization tasks, enabling direct generation of feasible structural layouts conditioned on performance targets~\cite{maze2023diffusion, keramati2026heatgen, kim2026human}. In this work, cooling-channel layouts are represented as binary flow masks, enabling generative modeling of channel topology. And the generated channels must satisfy physical and geometric constraints, including inlet-to-outlet connectivity, minimum channel width, symmetry, and coolant coverage over heat-generating chip regions.

We present a physics-guided generative design framework for liquid-cooling channel topology design in a high-power multi-chip electronic package. A conditional diffusion model is used to generate symmetric channel layouts, after which topology screening removes designs that lack a continuous inlet-to-outlet flow path or contain excessive dead-end branching. The remaining feasible designs are evaluated using a calibrated thermal-fluids reduced-order model based on three performance objectives: maximum GPU temperature ($T_{\max,\mathrm{GPU}}$), GPU temperature spread ($\Delta T_{\mathrm{GPU}}$), and pressure drop ($\Delta P$). Multi-objective analysis is then used to characterize the trade-off between thermal performance and hydraulic resistance and to identify high-performing candidate topologies. The selected design is subsequently reconstructed and independently evaluated using 3D conjugate heat-transfer simulations in OpenFOAM to assess the accuracy of the reduced-order thermal and hydraulic predictions. Overall, the proposed framework demonstrates that physics-guided generative design can discover non-conventional cooling-channel topologies that provide substantial improvements in both chip-level thermal performance and hydraulic resistance relative to a conventional reference design, while explicitly resolving the broader thermal-fluids trade-offs within the generated design space.

\section{Thermal-Fluids Model}
\label{sec:thermal_hydraulic_model}

\subsection{System Configuration and Design Objectives}
\label{subsec:system_configuration}

The thermal management problem considered in this study involves a high-power multi-chip electronic package with two GPU heat sources and one CPU heat source beneath a liquid-cooled cold plate. The package configuration is representative of modern high-density accelerator architectures, such as the NVIDIA GB200 multi-chip module, and is arranged within a rectangular domain of dimensions $L_x = 189.5~\mathrm{mm}$ and $L_y = 270.4~\mathrm{mm}$.

For reduced-order evaluation, the planform domain is discretized using a Cartesian grid of $N_x = N_y = 200$, giving spatial resolutions of $\Delta x = L_x/N_x$ and $\Delta y = L_y/N_y$. The same physical dimensions and component footprints are retained when selected channel geometries are reconstructed for 3D CFD validation.

The two GPUs are positioned symmetrically about the package centerline ($y = 130~\mathrm{mm}$), with a $43~\mathrm{mm}$ lateral separation between their adjacent inner edges. Each GPU has a footprint of $56.7 \times 52.8~\mathrm{mm}^2$ and dissipates $P_{\mathrm{GPU}} = 1{,}200~\mathrm{W}$. The CPU is centered laterally near the upper portion of the package ($y = 220~\mathrm{mm}$) and has a footprint of $22.8 \times 30.6~\mathrm{mm}^2$ with a heat dissipation of $P_{\mathrm{CPU}} = 300~\mathrm{W}$. The total applied heat load is given in Eq.~\eqref{eq:total_heat_load}.

\begin{equation}
    Q_{\mathrm{total}} = 2P_{\mathrm{GPU}} + P_{\mathrm{CPU}} = 2{,}700~\mathrm{W}
    \label{eq:total_heat_load}
\end{equation}

Heat generation is assumed to be spatially uniform within each chip footprint.
Cooling is provided through a liquid channel network located above the heat-generating substrate. A single $25.4~\mathrm{mm}$-wide inlet is positioned at the center of the lower boundary, with a corresponding outlet located at the center of the upper boundary. Single-phase liquid water enters at a fixed volumetric flow rate of $\dot{V} = 5~\mathrm{L\,min^{-1}}$ and an inlet temperature of $T_{\mathrm{in}} = 25^\circ\mathrm{C}$, with a channel depth of $H = 1~\mathrm{mm}$. All candidate designs operate under identical conditions, so differences in thermal and hydraulic performance arise solely from the channel topology.
Each cooling-channel topology is represented by a binary planform mask, as shown in Eq.~\eqref{eq:binary_mask}.

\begin{equation}
    \phi(x,y) =
    \begin{cases}
        1, & (x,y) \in \Omega_f \\
        0, & (x,y) \notin \Omega_f
    \end{cases}
    \label{eq:binary_mask}
\end{equation}
where $\Omega_f$ denotes the fluid region.
Generated layouts are constrained to use the prescribed inlet and outlet and to maintain left-right symmetry. A continuous inlet-to-outlet flow path is required before a design is retained for thermal-fluids evaluation.
The design problem is formulated using three thermal-fluids objectives. The first objective is the maximum temperature across the two GPU regions, as given by Eq.~\eqref{eq:tmax_gpu}.

\begin{equation}
    T_{\max,\mathrm{GPU}} = \max_{\mathbf{x}\in\Omega_{\mathrm{GPU}}} T(\mathbf{x})
    \label{eq:tmax_gpu}
\end{equation}
where $\Omega_{\mathrm{GPU}} = \Omega_{\mathrm{GPU,L}} \cup \Omega_{\mathrm{GPU,R}}$.
This metric characterizes local hot-spot suppression.
The second objective quantifies temperature nonuniformity across GPU regions, as given by Eq.~\eqref{eq:deltaT_gpu}.

\begin{equation}
    \Delta T_{\mathrm{GPU}} = T_{\max,\mathrm{GPU}} - T_{\min,\mathrm{GPU}}
    \label{eq:deltaT_gpu}
\end{equation}
where $T_{\min,\mathrm{GPU}} = \min_{\mathbf{x}\in\Omega_{\mathrm{GPU}}} T(\mathbf{x})$.
A lower value of $\Delta T_{\mathrm{GPU}}$ indicates a more spatially uniform GPU temperature distribution.
The third objective is the pressure drop between the inlet and outlet, as shown in Eq.~\eqref{eq:pressure_drop}.

\begin{equation}
    \Delta P = P_{\mathrm{in}} - P_{\mathrm{out}}
    \label{eq:pressure_drop}
\end{equation}
where $P_{\mathrm{in}}$ and $P_{\mathrm{out}}$ denote the inlet and outlet static pressures, respectively.
Because the volumetric flow rate is held constant for all candidate designs, minimizing $\Delta P$ directly minimizes the required pumping power.
The resulting design problem is formulated as a simultaneous multi-objective minimization problem shown in Eq.~\eqref{eq:design_objectives}.

\begin{equation}
    \min_{\phi} \left[ T_{\max,\mathrm{GPU}}(\phi), \; \Delta T_{\mathrm{GPU}}(\phi), \; \Delta P(\phi) \right]
    \label{eq:design_objectives}
\end{equation}

The minimization is subject to the prescribed package geometry, operating conditions, and topology-feasibility constraints.

\subsection{Reduced-Order Model}
\label{sec:rom}

Direct 3D conjugate heat-transfer simulation of every generated channel topology is computationally expensive during large scale exploration. A thermal-fluids reduced-order model (ROM) was therefore developed to evaluate candidate designs while capturing the dominant physical mechanisms governing in-plane heat spreading, depth-averaged flow distribution, local convective heat transfer, progressive coolant warming, and hydraulic resistance~\cite{lad2021reduced}. The model operates on the $200 \times 200$ Cartesian grid and evaluates only topologies satisfying through-flow connectivity.

The ROM adopts a 2.5D formulation. In-plane heat conduction through the solid stack is represented on the two-dimensional (2D) planform, while the finite thickness of the copper and silicon layers is incorporated through equivalent in-plane conductance and through-thickness thermal resistance. Coolant motion within the $1~\mathrm{mm}$ deep channel is solved using a conservative depth-averaged pressure-potential approximation, yielding a spatially varying convective heat-transfer coefficient. Coolant warming is calculated via an energy balance along the connected flow path, and a calibrated hydraulic-loss model provides the total inlet-to-outlet pressure drop.

\subsubsection{Equivalent Solid Conduction Model}

The solid stack comprises a $4~\mathrm{mm}$ copper cold-plate layer ($k_{\mathrm{Cu}} = 400~\mathrm{W\,m^{-1}K^{-1}}$) and a $2~\mathrm{mm}$ silicon device layer ($k_{\mathrm{Si}} = 150~\mathrm{W\,m^{-1}K^{-1}}$). An equivalent in-plane thermal conductivity is defined by preserving the total lateral conductance of the two layers, as shown in Eq.~\eqref{eq:k_effective}.

\begin{equation}
    k_{\mathrm{eff}} = \frac{k_{\mathrm{Cu}}t_{\mathrm{Cu}} + k_{\mathrm{Si}}t_{\mathrm{Si}}}{t_{\mathrm{Cu}} + t_{\mathrm{Si}}} = 316.7~\mathrm{W\,m^{-1}K^{-1}}
    \label{eq:k_effective}
\end{equation}

The corresponding total effective thickness is $t_{\mathrm{eff}} = t_{\mathrm{Cu}} + t_{\mathrm{Si}} = 6~\mathrm{mm}$. The steady planform temperature field is governed by Eq.~\eqref{eq:rom_energy}.

\begin{equation}
    -\nabla\cdot\left(k_{\mathrm{eff}}t_{\mathrm{eff}}\nabla T_p\right) + h_c(x,y)\left[T_p(x,y) - T_f(x,y)\right] = q''(x,y)
    \label{eq:rom_energy}
\end{equation}
where $T_p$ is the effective plate temperature, $h_c(x,y)$ is the local convective heat-transfer coefficient referenced to the planform area, $T_f(x,y)$ is the local coolant temperature, and $q''(x,y)$ is the prescribed chip heat-flux distribution.
Adiabatic conditions are imposed along the outer domain boundaries, as shown in Eq.~\eqref{eq:rom_adiabatic}.

\begin{equation}
    \mathbf{n}\cdot\nabla T_p = 0
    \label{eq:rom_adiabatic}
\end{equation}

Because the 2D conduction operator does not explicitly resolve the through-thickness temperature gradient within the stack, a series thermal resistance is applied over the heated chip footprints, as shown in Eq.~\eqref{eq:r_stack}.

\begin{equation}
    R_{\mathrm{stack}} = \frac{t_{\mathrm{Si}}}{k_{\mathrm{Si}}} + \frac{t_{\mathrm{Cu}}}{k_{\mathrm{Cu}}} = 2.33 \times 10^{-5}~\mathrm{m^2\,K\,W^{-1}}
    \label{eq:r_stack}
\end{equation}

The junction temperature field $T_j(x,y)$ used to evaluate hot-spot metrics is given by Eq.~\eqref{eq:junction_temperature}.

\begin{equation}
    T_j(x,y) = T_p(x,y) + q''(x,y)R_{\mathrm{stack}}
    \label{eq:junction_temperature}
\end{equation}

\subsubsection{Depth-Averaged Flow and Local Heat Transfer}

For each candidate topology, only the connected fluid component spanning both the inlet and outlet is retained. A conservative depth-averaged pressure potential field $p^*$ is solved exclusively within this fluid domain, as shown in Eq.~\eqref{eq:rom_pressure_potential}.

\begin{equation}
    \nabla^2 p^* = 0 \quad \mathrm{in}\ \Omega_f
    \label{eq:rom_pressure_potential}
\end{equation}

The solution is subject to Dirichlet boundary conditions $p^* = 1$ at the inlet and $p^* = 0$ at the outlet, with no-flux conditions along solid channel walls ($\mathbf{n}\cdot\nabla p^* = 0$).
The unscaled depth-averaged velocity field $\mathbf{u}^* = -\nabla p^*$ is scaled to strictly enforce the prescribed volumetric flow rate $\dot{V} = 5~\mathrm{L\,min^{-1}}$, as shown in Eq.~\eqref{eq:flow_scaling}.

\begin{equation}
    \mathbf{u} = S_Q\,\mathbf{u}^*, \qquad S_Q = \frac{\dot{V}}{\dot{V}^*}
    \label{eq:flow_scaling}
\end{equation}
where $\dot{V}^*$ is the integrated inlet flux of the normalized solution.
The local channel width is estimated from the Euclidean wall-distance transform, $w(x,y) = 2d_w(x,y)$. The local hydraulic diameter for channel depth $H = 1~\mathrm{mm}$ is given by Eq.~\eqref{eq:hydraulic_diameter}.

\begin{equation}
    D_h(x,y) = \frac{2w(x,y)H}{w(x,y) + H}
    \label{eq:hydraulic_diameter}
\end{equation}

The local Reynolds number is calculated by $Re(x,y) = \rho_f |\mathbf{u}(x,y)| D_h(x,y) / \mu_f$. In the laminar regime, a fully developed Nusselt number $Nu_{\mathrm{lam}} = 5.39$ is assigned~\cite{shah1978laminar}. In the turbulent regime, the Gnielinski correlation is employed~\cite{gnielinski1976new}, as shown in Eq.~\eqref{eq:gnielinski}.

\begin{equation}
    Nu_{\mathrm{turb}} = \frac{(f/8)(Re-1000)Pr}{1 + 12.7(f/8)^{1/2}\left(Pr^{2/3}-1\right)}
    \label{eq:gnielinski}
\end{equation}

The Darcy friction factor is $f = [0.79\ln(Re) - 1.64]^{-2}$. A linear transition is applied between $Re = 2{,}000$ and $Re = 4{,}000$~\cite{joseph2010friction}, as shown in Eq.~\eqref{eq:nu_transition}.

\begin{equation}
    Nu = (1-\gamma)Nu_{\mathrm{lam}} + \gamma Nu_{\mathrm{turb}}, \quad \gamma = \mathrm{clip}\left(\frac{Re-2000}{2000}, \, 0, \, 1\right)
    \label{eq:nu_transition}
\end{equation}

The local convective heat-transfer coefficient is calculated using Eq.~\eqref{eq:rom_htc}.

\begin{equation}
    h_c(x,y) = C_h \frac{Nu(x,y)k_f}{D_h(x,y)}
    \label{eq:rom_htc}
\end{equation}
where $k_f$ is the coolant thermal conductivity and $C_h = 2.6841$ is a calibrated thermal enhancement factor.

\subsubsection{Path-Dependent Coolant Warming}

Coolant temperature is updated along the channel to account for stream-wise heat absorption. The normalized path progress variable $s(x,y)$ is defined in Eq.~\eqref{eq:path_progress}.

\begin{equation}
    s(x,y) = \frac{d_f(x,y)}{d_{f,\max}}, \qquad 0 \le s \le 1
    \label{eq:path_progress}
\end{equation}
where $d_f(x,y)$ is the shortest path distance from the inlet through connected fluid cells and $d_{f,\max}$ is the maximum path length.
For an absorbed heat rate $\dot{Q}_{\mathrm{abs}}$, the bulk coolant temperature rise is $\Delta T_f = \dot{Q}_{\mathrm{abs}} / (\dot{m} c_p)$. The resulting local coolant temperature distribution is given by Eq.~\eqref{eq:path_coolant_temperature}.

\begin{equation}
    T_f(x,y) = T_{\mathrm{in}} + s(x,y)\Delta T_f
    \label{eq:path_coolant_temperature}
\end{equation}

Coupling between the solid conduction and coolant advection is resolved iteratively. After each solve of Eq.~\eqref{eq:rom_energy}, the absorbed heat is updated using Eq.~\eqref{eq:q_abs_rom}.

\begin{equation}
    \dot{Q}_{\mathrm{abs}} = \int_{\Omega_f} h_c(x,y) \max\left[T_p(x,y) - T_f(x,y), \, 0\right] dA
    \label{eq:q_abs_rom}
\end{equation}

The coolant temperature $T_f(x,y)$ is refreshed using Eq.~\eqref{eq:path_progress} and Eq.~\eqref{eq:path_coolant_temperature}. Four outer iterations are used to ensure stable thermal convergence.

\subsubsection{Hydraulic Pressure-Drop Model}
\label{sec:rom_pressure_drop}

The pressure drop is evaluated using a 1D loss accumulation model. At each streamwise row $j$, the total fluid width is $w_j = N_{f,j}\Delta x$, where $N_{f,j}$ is the row fluid cell count. The corresponding flow area, mean velocity, and hydraulic diameter are $A_j = w_j H$, $V_j = \dot{m} / (\rho_f A_j)$, and $D_{h,j} = 4A_j / [2(w_j+H)]$, respectively.
The Darcy friction factor is computed using Eq.~\eqref{eq:rom_friction_factor}.

\begin{equation}
    f_j =
    \begin{cases}
        \dfrac{64}{Re_j}, & Re_j < 2{,}300 \\[6pt]
        0.316\,Re_j^{-0.25}, & Re_j \ge 2{,}300
    \end{cases}
    \label{eq:rom_friction_factor}
\end{equation}

Channel tortuosity is defined as $\tau = L_{\mathrm{path}} / L_y$, where $L_{\mathrm{path}}$ is the maximum connected inlet-to-outlet path length. The frictional pressure loss is accumulated according to Eq.~\eqref{eq:rom_dp_friction}.

\begin{equation}
    \Delta P_{\mathrm{fric}} = \sum_j f_j \frac{\tau\Delta y}{D_{h,j}} \frac{\rho_f V_j^2}{2}
    \label{eq:rom_dp_friction}
\end{equation}

Sudden expansion and contraction losses between adjacent rows ($K_{\mathrm{exp}} = [1 - (A_{j-1}/A_j)]^2$ for $A_j > A_{j-1}$, and $K_{\mathrm{con}} = \frac{1}{2}[1 - (A_j/A_{j-1})]$ for $A_j < A_{j-1}$) are accumulated as $\Delta P_{\mathrm{minor}} = \sum_j K_j \rho_f V_{\mathrm{ref},j}^2 / 2$.
The total ROM pressure drop is given by Eq.~\eqref{eq:rom_dp_total}.

\begin{equation}
    \Delta P_{\mathrm{ROM}} = C_p \left(\Delta P_{\mathrm{fric}} + \Delta P_{\mathrm{minor}}\right)
    \label{eq:rom_dp_total}
\end{equation}
where $C_p = 1.89$ is the hydraulic calibration coefficient.
The complete set of parameters used in the ROM is summarized in Table~\ref{tab:rom_parameters}.

\begin{table}[h]
\centering
\caption{Principal parameters used in the reduced-order thermal-fluids model.}
\label{tab:rom_parameters}
\begin{tabular}{lll}
\hline
Parameter & Symbol & Value \\
\hline
Copper conductivity & $k_{\mathrm{Cu}}$ & $400~\mathrm{W\,m^{-1}K^{-1}}$ \\
Copper thickness & $t_{\mathrm{Cu}}$ & $4~\mathrm{mm}$ \\
Silicon conductivity & $k_{\mathrm{Si}}$ & $150~\mathrm{W\,m^{-1}K^{-1}}$ \\
Silicon thickness & $t_{\mathrm{Si}}$ & $2~\mathrm{mm}$ \\
Equivalent in-plane conductivity & $k_{\mathrm{eff}}$ & $316.7~\mathrm{W\,m^{-1}K^{-1}}$ \\
Channel depth & $H$ & $1~\mathrm{mm}$ \\
Coolant density & $\rho_f$ & $997~\mathrm{kg\,m^{-3}}$ \\
Coolant specific heat & $c_p$ & $4181~\mathrm{J\,kg^{-1}K^{-1}}$ \\
Dynamic viscosity & $\mu_f$ & $8.9\times 10^{-4}~\mathrm{Pa\,s}$ \\
Prandtl number & $Pr$ & $6.2$ \\
Coolant conductivity & $k_f$ & $0.600~\mathrm{W\,m^{-1}K^{-1}}$ \\
Thermal calibration coefficient & $C_h$ & $2.6841$ \\
Hydraulic calibration coefficient & $C_p$ & $1.89$ \\
\hline
\end{tabular}
\end{table}

\subsubsection{ROM Calibration}
\label{sec:rom_calibration}

The reduced-order formulation contains deliberate simplifications relative to the 3D conjugate heat-transfer model, including depth-averaged flow, 2D heat spreading, correlation-based convection, and lumped hydraulic losses. Two global calibration coefficients are therefore retained in the final ROM: the thermal coefficient $C_h$, which scales the local Nusselt-number-based heat-transfer coefficient in Eq.~\eqref{eq:rom_htc}, and the hydraulic coefficient $C_p$, which scales the accumulated frictional and minor pressure losses in Eq.~\eqref{eq:rom_dp_total}. These coefficients are applied globally rather than being adjusted for individual generated topologies.

The thermal calibration was performed using the conservative fluid-only depth-averaged flow solver described previously, in which the velocity field is restricted to the connected channel region, and exact no-flow conditions are imposed at fluid--solid boundaries. A single shared heat-transfer multiplier was fitted across the thermal calibration cases, giving $C_h = 2.6841$. The calibration accuracy was assessed using the estimated temperature rise above the coolant inlet temperature. For a reference temperature $T_{\mathrm{ref}}$, the relative temperature-rise error is defined as shown in Eq.~\eqref{eq:thermal_calibration_error}.

\begin{equation}
    \varepsilon_{\Delta T}
    =
    \frac{
        \left|
        (T_{\mathrm{ROM}}-T_{\mathrm{in}})
        -
        (T_{\mathrm{ref}}-T_{\mathrm{in}})
        \right|
    }{
        T_{\mathrm{ref}}-T_{\mathrm{in}}
    }
    \times100\%
    \label{eq:thermal_calibration_error}
\end{equation}

The fitted value of $C_h$ produced a mean temperature-rise error of $7.7\%$ and a maximum error of $12.6\%$ across the calibration cases.
Following calibration, $C_h$ and $C_p$ are held fixed for all subsequent topology evaluations. No design-specific adjustment of either coefficient is performed during screening, multi-objective ranking, or final candidate assessment. The purpose of the ROM is therefore not to reproduce every local feature of the 3D flow and temperature fields, but to provide a computationally efficient and physically consistent estimate of the relative thermal-fluids performance of large populations of candidate channel topologies. Independent high-fidelity CFD validation of the final selected design is presented subsequently to assess the predictive accuracy of the calibrated ROM outside the calibration procedure.

\subsection{Full-Order Conjugate Heat-Transfer Model}
\label{sec:full_order_cfd}

A 3D conjugate heat-transfer (CHT) model was developed in OpenFOAM to provide an independent high-fidelity assessment of the final generated cooling channel topology. The 2D channel mask selected from the generative design workflow was reconstructed as a 3D fluid domain by extrusion through the prescribed $1~\mathrm{mm}$ channel depth. The complete computational model consists of three coupled regions: the liquid-coolant region, a copper cold-plate region, and a heat-generating device layer representing the GPU and CPU regions. The three regions are solved simultaneously so that the fluid flow, convective heat transfer, solid conduction, and temperature continuity across the material interfaces are resolved directly.

The full-order simulations were performed using OpenFOAM Foundation version 13 with a multi-region finite-volume formulation~\cite{weller1998tensorial}. The coolant is modeled as an incompressible Newtonian fluid with constant thermophysical properties. Because the generated topology contains strong contractions, expansions, and local recirculation regions, turbulence effects are represented using the Reynolds-averaged Navier--Stokes (RANS) equations with the $k$--$\omega$ shear-stress-transport (SST) model~\cite{menter1994two}. The solid copper and device regions are modeled using isotropic heat conduction with constant material properties.

\subsubsection{Governing Equations}
\label{sec:cfd_governing}

For steady incompressible flow, conservation of mass is expressed as
\begin{equation}
    \nabla\cdot\mathbf{u}=0
    \label{eq:cfd_continuity}
\end{equation}
where $\mathbf{u}$ is the mean coolant velocity vector.
The Reynolds-averaged momentum equation is written in Eq.~\eqref{eq:cfd_momentum}.

\begin{equation}
    \rho_f (\mathbf{u}\cdot\nabla)\mathbf{u} = -\nabla p + \nabla\cdot\left[\mu_{\mathrm{eff}}\left(\nabla\mathbf{u} + \nabla\mathbf{u}^{T}\right)\right]
    \label{eq:cfd_momentum}
\end{equation}
where $\rho_f$ is the coolant density, $p$ is pressure, and $\mu_{\mathrm{eff}}$ is the effective viscosity.

\begin{equation}
    \mu_{\mathrm{eff}} = \mu_f+\mu_t
\end{equation}

It includes the molecular viscosity $\mu_f$ and turbulent eddy viscosity $\mu_t$ estimated by the $k$--$\omega$ SST turbulence model.
The steady fluid-energy equation is expressed as
\begin{equation}
    \rho_f c_{p,f}\mathbf{u}\cdot\nabla T_f = \nabla\cdot\left(k_{f,\mathrm{eff}}\nabla T_f\right)
    \label{eq:cfd_fluid_energy}
\end{equation}
where $T_f$ is the coolant temperature and $k_{f,\mathrm{eff}}$ includes both molecular and turbulent thermal transport. Heat transfer in each solid region is governed by
\begin{equation}
    \nabla\cdot\left(k_s\nabla T_s\right) + q'''=0
    \label{eq:cfd_solid_energy}
\end{equation}
where $k_s$ is the solid thermal conductivity and $q'''$ is the volumetric heat generation term. In the copper cold plate, $q'''=0$, whereas heat generation is applied within the GPU and CPU cell zones of the device layer.
At every fluid--solid and solid--solid interface, continuity of temperature and normal heat flux is enforced.
\begin{equation}
    T_1=T_2
    \label{eq:interface_temperature}
\end{equation}

\begin{equation}
    -k_1\frac{\partial T_1}{\partial n} = -k_2\frac{\partial T_2}{\partial n}
    \label{eq:interface_heatflux}
\end{equation}

The conditions shown in Eq.~\eqref{eq:interface_temperature} and Eq.~\eqref{eq:interface_heatflux} couple coolant convection, copper heat spreading, and device-layer conduction without prescribing an empirical heat-transfer coefficient at the interface.

\subsubsection{Boundary Conditions and Material Properties}
\label{sec:cfd_boundary_conditions}

The CFD model uses the same package geometry, heat loads, coolant flow rate, and inlet temperature as the reduced-order model so that the two approaches can be compared under matched operating conditions. Liquid water enters through the lower inlet at a volumetric flow rate of $\dot{V}=5~\mathrm{L\,min^{-1}}$, and a fixed temperature of $T_{\mathrm{in}}=25^\circ\mathrm{C} = 298.15~\mathrm{K}$.

The corresponding inlet velocity is prescribed to produce the target mass flow rate. A reference pressure is imposed at the outlet, while the velocity satisfies the appropriate outflow condition. No-slip velocity conditions are imposed along all coolant channel walls.

The outlet temperature uses an inlet--outlet formulation: a zero-gradient condition applies during outward flow, while $T_{\mathrm{in}}=298.15~\mathrm{K}$ is imposed if local reverse flow occurs. The outer channel walls are treated as adiabatic,
\begin{equation}
    \mathbf{n}\cdot\nabla T=0
\end{equation}
while the fluid-copper and copper-device interfaces use coupled temperature boundary conditions. External surfaces of the solid regions are also treated as adiabatic so that the applied chip power is removed through the liquid-cooling interface.

The heat-generating device layer is divided into three cell zones corresponding to the two GPUs and the CPU. Mesh-independent total-power source terms are applied to these zones:
\begin{equation}
    Q_{\mathrm{GPU,L}}=1{,}200~\mathrm{W} \quad Q_{\mathrm{GPU,R}}=1{,}200~\mathrm{W} \quad Q_{\mathrm{CPU}}=300~\mathrm{W}
\end{equation}

Thus, the total heat-generation rate can be calculated, $Q_{\mathrm{total}}=2700~\mathrm{W}$. The thermophysical properties used in the full order CHT model are summarized in Table~\ref{tab:cfd_properties}.

\begin{table}[htbp]
\centering
\caption{Thermophysical properties used in the full-order CHT model.}
\label{tab:cfd_properties}

\begin{tabular}{lccc}
\hline
Property & Water & Copper & Silicon/device layer \\
\hline
Density, $\rho$ [$\mathrm{kg\,m^{-3}}$] & 997 & 8960 & 2330 \\
Specific heat, $c_p$ [$\mathrm{J\,kg^{-1}K^{-1}}$] & 4181 & 385 & 700 \\
Thermal conductivity, $k$ [$\mathrm{W\,m^{-1}K^{-1}}$] & 0.600 & 400 & 150 \\
Dynamic viscosity, $\mu$ [$\mathrm{Pa\,s}$] & $8.9\times10^{-4}$ & -- & -- \\
\hline
\end{tabular}
\end{table}

Fluid and solid properties are assumed constant over the temperature range considered, and gravitational body forces are neglected.

\subsubsection{Geometry Reconstruction and Mesh}
\label{sec:cfd_mesh}

The selected binary channel mask is converted into a 3D coolant geometry while preserving the planform topology used by the reduced-order model. The coolant region is extruded through a channel depth of $1~\mathrm{mm}$ and positioned above the copper cold plate. Separate volumetric meshes are generated for the fluid, copper, and heat-generating device regions. The copper and device meshes retain the chip locations and thicknesses used in the reduced-order formulation.

The three region meshes are connected using non-conformal mapped interfaces. The fluid-to-copper and copper-to-device interfaces are paired so that temperature and heat flux are transferred conservatively between adjacent regions. Mesh connectivity and interface coverage are checked before each simulation to ensure that no uncoupled interface faces remain.

Mesh quality is evaluated using OpenFOAM's standard mesh-quality diagnostics, including cell volume, non-orthogonality, skewness, aspect ratio, and region connectivity. The channel depth is discretized into multiple cells to resolve velocity and temperature gradients normal to the cooled surface. 

\subsubsection{Numerical Solution and Convergence}
\label{sec:cfd_convergence}

The governing equations are discretized using the finite-volume method and solved in steady-state mode. Pressure--velocity coupling is handled using the PIMPLE framework operated with a single outer corrector, which reduces to a SIMPLE-type steady-state solution procedure~\cite{issa1986solution, patankar1983calculation}. The fluid energy equation and the solid energy equations are solved simultaneously with the momentum and turbulence equations through the multi-region coupling procedure.

\begin{figure}[!t]
    \centering
    \includegraphics[
        width=\textwidth,
        height=0.76\textheight,
        keepaspectratio
    ]{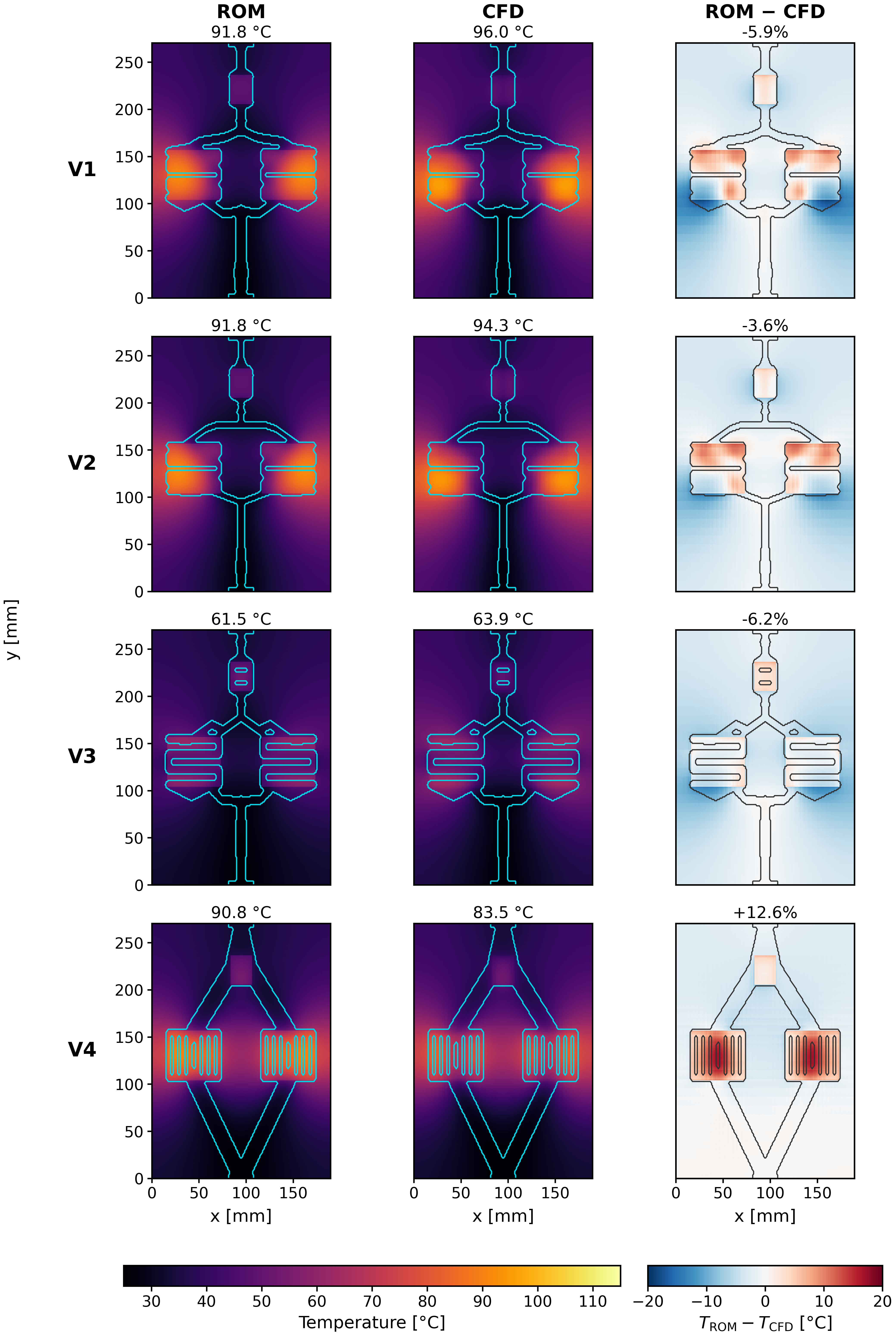}
    \caption{Comparison of reduced-order model and CFD junction-temperature
    predictions for 4 representative cases.}
    \label{fig:rom_cfd_validation}
\end{figure}

Convergence is monitored using both algebraic residuals and physically conserved quantities. The residual tolerances employed in the final simulations are $R_U \le 10^{-5}$, $R_h \le 10^{-5}$, $R_{p} \le 10^{-4}$, $R_k \le 10^{-4}$, $R_{\omega} \le 10^{-4}$ for the fluid region, and
$R_e \le 10^{-6}$ for each solid region.

Residual convergence alone is not used as the sole acceptance criterion. Mass conservation is monitored using the integrated inlet and outlet mass-flow rates, while energy conservation is assessed using the mixing-cup outlet temperature and the integrated heat flux through the coupled fluid--solid interface. The coolant-side energy balance is evaluated from Eq.~\eqref{eq:cfd_energy_balance}.

\begin{equation}
    \dot{Q}_f = \dot{m}c_p\left(T_{\mathrm{out,mw}} - T_{\mathrm{in}}\right)
    \label{eq:cfd_energy_balance}
\end{equation}
where $T_{\mathrm{out,mw}}$ is the mass-flow-weighted outlet temperature.
A solution is considered converged when the monitored temperatures and pressure drop become statistically stationary, inlet and outlet mass flow rates agree, the prescribed residual thresholds are satisfied, and the integrated coolant heat removal approaches the applied $2{,}700~\mathrm{W}$ heat load. The resulting converged fields are then post-processed to obtain $T_{\max,\mathrm{GPU}}$, $\Delta T_{\mathrm{GPU}}$, $\Delta P$, outlet temperature, and the overall energy balance.

\subsection{Reduced-Order Model Validation}
\label{sec:rom_validation}

The reduced-order model was assessed against 3D OpenFOAM conjugate heat-transfer simulations before being used to evaluate large-scale channel topologies. Representative channel layouts exhibiting different flow-path characteristics were reconstructed from the same binary masks used by the ROM and simulated under matched geometric, thermal, and hydraulic operating conditions. In both models, the coolant flow rate was fixed at $5~\mathrm{L\,min^{-1}}$, the inlet temperature was $25^\circ\mathrm{C}$, and the total applied heat load was $2{,}700~\mathrm{W}$. The same chip footprints, material properties, and channel depth were retained so that differences between the two models could be attributed primarily to the reduced-order approximations.

Validation focuses on the quantities most directly related to the design objectives: peak GPU temperature and inlet-to-outlet pressure drop. Temperature-field distributions are also examined qualitatively to determine whether the ROM reproduces the location and general spatial structure of the dominant hot regions. Because the ROM and CFD simulations use the same coolant inlet temperature, thermal error is evaluated as the temperature rise above the inlet rather than as an absolute temperature. This avoids artificially reducing the relative error by using a common temperature offset.

For the maximum GPU temperature, the relative temperature-rise error is defined in Eq.~\eqref{eq:validation_tmax_error}.
\begin{equation}
    \varepsilon_{T_{\max}}
    =
    \frac{
        \left|
        \left(T_{\max,\mathrm{GPU}}^{\mathrm{ROM}}-T_{\mathrm{in}}\right)
        -
        \left(T_{\max,\mathrm{GPU}}^{\mathrm{CFD}}-T_{\mathrm{in}}\right)
        \right|
    }{
        T_{\max,\mathrm{GPU}}^{\mathrm{CFD}}-T_{\mathrm{in}}
    }
    \times100\%
    \label{eq:validation_tmax_error}
\end{equation}

The hydraulic prediction is evaluated using the relative pressure-drop error, as shown in Eq. \eqref{eq:validation_dp_error}.

\begin{equation}
    \varepsilon_{\Delta P}
    =
    \frac{
        \left|
        \Delta P_{\mathrm{ROM}}
        -
        \Delta P_{\mathrm{CFD}}
        \right|
    }{
        \Delta P_{\mathrm{CFD}}
    }
    \times100\%
    \label{eq:validation_dp_error}
\end{equation}

For GPU temperature nonuniformity, the corresponding relative error can be calculated through Eq.~\eqref{eq:validation_deltaT_error}.

\begin{equation}
    \varepsilon_{\Delta T}
    =
    \frac{
        \left|
        \Delta T_{\mathrm{GPU}}^{\mathrm{ROM}}
        -
        \Delta T_{\mathrm{GPU}}^{\mathrm{CFD}}
        \right|
    }{
        \Delta T_{\mathrm{GPU}}^{\mathrm{CFD}}
    }
    \times100\%
    \label{eq:validation_deltaT_error}
\end{equation}

Figure~\ref{fig:rom_cfd_validation} shows the comparison of the spatial temperature estimation of the reduced-order and full-order models. The ROM reproduces the principal hotspot locations and the overall temperature distribution across the representative channel topologies.
In addition to the scalar quantities, the ROM's spatial thermal behavior is examined against the corresponding CFD temperature distribution. Because the CFD model resolves the complete 3D solid stack whereas the ROM predicts an equivalent 2.5D junction-temperature field, the comparison is performed on a consistent projected chip-level surface. A local temperature-difference field may be expressed in Eq.~\eqref{eq:validation_field_error}.

\begin{equation}
    E_T(x,y)
    =
    T_{\mathrm{ROM}}(x,y)
    -
    T_{\mathrm{CFD}}(x,y)
    \label{eq:validation_field_error}
\end{equation}
where the sign of $E_T$ indicates whether the reduced-order formulation locally overpredicts or underpredicts the CFD temperature.

The CFD reference solutions are accepted only after mass flow conservation, thermal convergence, and energy conservation are established. The coolant-side heat-removal rate is evaluated using Eq.~\eqref{eq:validation_energy_balance}.
\begin{equation}
    \dot{Q}_{f}
    =
    \dot{m}c_p
    \left(
        T_{\mathrm{out,mw}}-T_{\mathrm{in}}
    \right)
    \label{eq:validation_energy_balance}
\end{equation}

The use of the mass flow-weighted outlet temperature ensures consistency with the energy balance and avoids biases associated with a simple area-averaged outlet temperature.
The purpose of this validation stage is to establish that the calibrated ROM reproduces the dominant thermal and hydraulic behavior with sufficient accuracy for comparative design screening. The ROM is therefore used to evaluate the large design population generated, while the full 3D CHT model is reserved for selected high-fidelity checks. The final topology identified by the generative design and multi-objective ranking procedure is evaluated independently in Section~\ref{sec:selected_design_validation}, providing an additional out-of-sample assessment of the reduced-order predictions.

\section{Dataset Generation and Conditional Diffusion Model}
\label{sec:dataset_generation_generative_model}

\subsection{Physics-Guided Training Dataset}
\label{subsec:dataset_generation}

A physics-guided dataset was constructed to provide the conditional diffusion model with a diverse set of physically meaningful liquid-cooling channel topologies. Each design is represented as a binary flow mask on the same $200\times200$ structured Cartesian grid used by the reduced-order thermal-fluids model. The channel indicator is defined in Eq.~\eqref{eq:dataset_binary_mask}.

\begin{equation}
    \phi(x,y)
    =
    \begin{cases}
        1, & (x,y)\in\Omega_f \\
        0, & (x,y)\in\Omega_s
    \end{cases}
    \label{eq:dataset_binary_mask}
\end{equation}
where $\Omega_f$ and $\Omega_s$ denote the fluid and solid regions, respectively. This image-based representation allows the channel design problem to be treated using generative image modeling while retaining a direct correspondence between the generated pixels and the physical coolant domain.

The initial design library contains 5{,}000 channel layouts generated using stochastic geometric construction procedures that introduce variations in channel branching, flow-path placement, chip coverage, and overall fluid fraction. The generation procedure was designed to encourage coolant paths that originate from the prescribed bottom-center inlet, spread toward the heat-generating regions, and reconnect toward the top-center outlet. Because the package and thermal loading are symmetric about the vertical centerline, the training geometries were constructed to preserve left-right symmetry.

Several physical and geometric constraints were imposed before a topology was admitted to the final conditional-training set. Each retained design was required to contain a continuous coolant path connecting the inlet and outlet, while disconnected fluid components were removed. Designs with excessively small channel features or unrealistic fluid fractions were excluded. Coolant coverage over the GPU and CPU footprints was also considered so that the training distribution favored layouts in which coolant reaches the principal heat-generating regions rather than bypassing them.

Topology quality was additionally characterized using the dead-end branch ratio, as shown in Eq.~\eqref{eq:branch_ratio_screening}.

\begin{equation}
    F_{\mathrm{branch}}
    =
    \frac{L_{\mathrm{dead}}}{L_{\mathrm{skel}}}
    \label{eq:branch_ratio_training}
\end{equation}
where $L_{\mathrm{dead}}$ is the total skeleton length associated with dead-end branches and $L_{\mathrm{skel}}$ is the total channel-skeleton length. This descriptor was used during dataset preparation and model training to place greater emphasis on topologies containing clean through-flow pathways while retaining sufficient geometric diversity for generative learning.

After topology preprocessing, 3{,}213 layouts were retained for the final three-objective conditional-diffusion dataset. Each retained topology was associated with the thermal-fluids quantities used in the final design problem: maximum GPU temperature ($T_{\max,\mathrm{GPU}}$), GPU temperature spread ($\Delta T_{\mathrm{GPU}}$), and inlet-to-outlet pressure drop ($\Delta P$). These quantities were obtained from the reduced-order thermal-fluids framework described in Section~\ref{sec:rom}.

For the generative model, the corresponding performance-conditioning vector can be calculated through Eq.~\eqref{eq:diffusion_condition_vector}.
\begin{equation}
    \mathbf{c}
    =
    \left[
        \widetilde{T}_{\max,\mathrm{GPU}}, \,
        \widetilde{\Delta T}_{\mathrm{GPU}}, \,
        \widetilde{\log(\Delta P)}
    \right]
    \label{eq:diffusion_condition_vector}
\end{equation}
where the tilde denotes normalization using statistics from the training set. A logarithmic transformation is applied to $\Delta P$ before normalization because the hydraulic values span a substantially wider numerical range than the temperature quantities. The resulting condition vector allows the generative model to learn relationships between channel geometry and the three thermal-fluids performance objectives.

Because the final generator operates on a half-domain representation, each symmetric $200\times200$ training mask is reduced to a $200\times100$ half-domain image. The complete topology can subsequently be recovered by reflection about the package centerline. This reduces the generative search space and enforces the required geometric symmetry directly rather than relying on the neural network to learn symmetry from the data alone.

\subsection{Conditional Diffusion Model}
\label{subsec:conditional_diffusion}

A conditional denoising diffusion model was developed to generate new cooling-channel topologies from specified thermal-fluids performance targets. Diffusion probabilistic models learn a data distribution by gradually corrupting training samples with Gaussian noise and training a neural network to reverse this process~\cite{ho2020denoising}. A U-Net architecture is employed as the noise-prediction network because its multiscale encoder-decoder structure is well suited to spatially structured image-generation tasks~\cite{ronneberger2015u}.

Let $\mathbf{x}_0$ denote a clean binary half-domain channel mask. During the forward diffusion process, Gaussian noise is progressively added according to Eq.~\eqref{eq:forward_diffusion}.

\begin{equation}
    q(\mathbf{x}_t|\mathbf{x}_{t-1})
    =
    \mathcal{N}
    \left(
        \sqrt{\alpha_t}\mathbf{x}_{t-1}, \,
        (1-\alpha_t)\mathbf{I}
    \right)
    \label{eq:forward_diffusion}
\end{equation}
where $t$ denotes the diffusion time step and $\alpha_t=1-\beta_t$, with $\beta_t$ controlling the prescribed noise schedule. The cumulative product of $\alpha_t$ is defined in Eq.~\eqref{eq:alpha_bar}.
\begin{equation}
    \bar{\alpha}_t
    =
    \prod_{s=1}^{t}\alpha_s
    \label{eq:alpha_bar}
\end{equation}

A noisy sample can be generated directly from the clean topology using Eq.~\eqref{eq:diffusion_closed_form}.
\begin{equation}
    \mathbf{x}_t
    =
    \sqrt{\bar{\alpha}_t}\mathbf{x}_0
    +
    \sqrt{1-\bar{\alpha}_t}\boldsymbol{\epsilon},
    \qquad
    \boldsymbol{\epsilon}\sim\mathcal{N}(\mathbf{0},\mathbf{I})
    \label{eq:diffusion_closed_form}
\end{equation}

As $t$ increases, the original channel structure is progressively destroyed until the sample approaches Gaussian noise.

\subsubsection{Conditional U-Net Denoiser}
\label{subsec:conditional_unet}

The reverse process is represented using a conditional U-Net noise predictor,
\begin{equation}
    \boldsymbol{\epsilon}_{\theta}
    =
    \boldsymbol{\epsilon}_{\theta}
    \left(
        \mathbf{x}_t, t, \mathbf{c}
    \right)
    \label{eq:conditional_noise_predictor}
\end{equation}
where $\theta$ denotes the trainable network parameters and $\mathbf{c}$ is the performance-conditioning vector defined in Eq.~\eqref{eq:diffusion_condition_vector}. The diffusion time step and performance conditions are embedded in the residual blocks of the denoising network, so the reconstruction process is influenced by both the current noise level and the desired thermal-fluids behavior.
The model is trained using the standard noise-prediction objective, as shown in Eq.~\eqref{eq:diffusion_training_loss}.

\begin{equation}
    \mathcal{L}_{\mathrm{diff}}
    =
    \mathbb{E}_{\mathbf{x}_0,t,\boldsymbol{\epsilon}}
    \left[
        w_i
        \left\|
            \boldsymbol{\epsilon}
            -
            \boldsymbol{\epsilon}_{\theta}
            \left(
                \mathbf{x}_t,t,\mathbf{c}
            \right)
        \right\|_2^2
    \right]
    \label{eq:diffusion_training_loss}
\end{equation}

The topology-dependent sample weight $w_i$ is computed from the dead-end branch ratio as shown in Eq.~\eqref{eq:topology_weight}.

\begin{equation}
    w_i
    =
    \mathrm{clip}
    \left[
        \exp
        \left(
            -3\frac{F_{\mathrm{branch},i}}
                    {F_{\mathrm{limit}}}
        \right),
        0.25,
        1.0
    \right]
    \label{eq:topology_weight}
\end{equation}
where $F_{\mathrm{branch},i}$ is the dead-end branch ratio of training sample $i$ and $F_{\mathrm{limit}}$ is the prescribed branch-ratio reference value. Consequently, branch-free and low-branching topologies receive larger training weights, while designs containing greater dead-end branching retain a minimum weight of $0.25$ rather than being completely excluded from training.

Classifier-free conditioning is incorporated by randomly removing the performance condition during a portion of the training updates. A condition-presence indicator is therefore included internally in the conditioning embedding to distinguish conditional and unconditional samples. This enables the same network to estimate both conditional and unconditional noise during sampling~\cite{ho2022classifier}.

\subsubsection{Transfer Learning and Training Procedure}
\label{subsec:diffusion_training}

The final three-objective model was obtained using transfer learning from a previously converged conditional diffusion checkpoint. The earlier model was conditioned using maximum temperature and pressure-drop information. For the final formulation, the condition representation was extended to include GPU temperature spread, $\Delta T_{\mathrm{GPU}}$, resulting in the three final physical targets $T_{\max,\mathrm{GPU}}$, $\Delta T_{\mathrm{GPU}}$, and $\Delta P$.

Transfer learning was performed in two stages. First, the newly introduced temperature-spread conditioning pathway was trained while the previously learned topology-generation features were retained. The condition-embedding and associated residual-conditioning projections were subsequently fine-tuned so that the network could adapt to the complete three-objective conditioning space without destroying the geometric information learned during earlier training.

The final conditional dataset was divided into 2{,}909 training samples and 304 validation samples. Training employed a linear variance schedule with $\beta_t$ increasing from $1\times10^{-4}$ to $2\times10^{-2}$ over 200 diffusion steps. Validation loss was monitored during transfer learning, and the best-performing model checkpoint was retained for subsequent topology generation.

\subsubsection{Half-Domain Generation and Symmetry Reconstruction}
\label{subsec:half_domain_generation}

New channel designs are generated from Gaussian noise in the $200\times100$ half-domain rather than over the complete package. For a specified condition vector $\mathbf{c}$, the sampling process is shown in Eq.~\eqref{eq:rd}.

\begin{equation}
    \mathbf{x}_T \sim \mathcal{N}(\mathbf{0},\mathbf{I})
    \label{eq:rd}
\end{equation}

\begin{figure}[t]
    \centering
    \includegraphics[
        width=0.98\textwidth
    ]{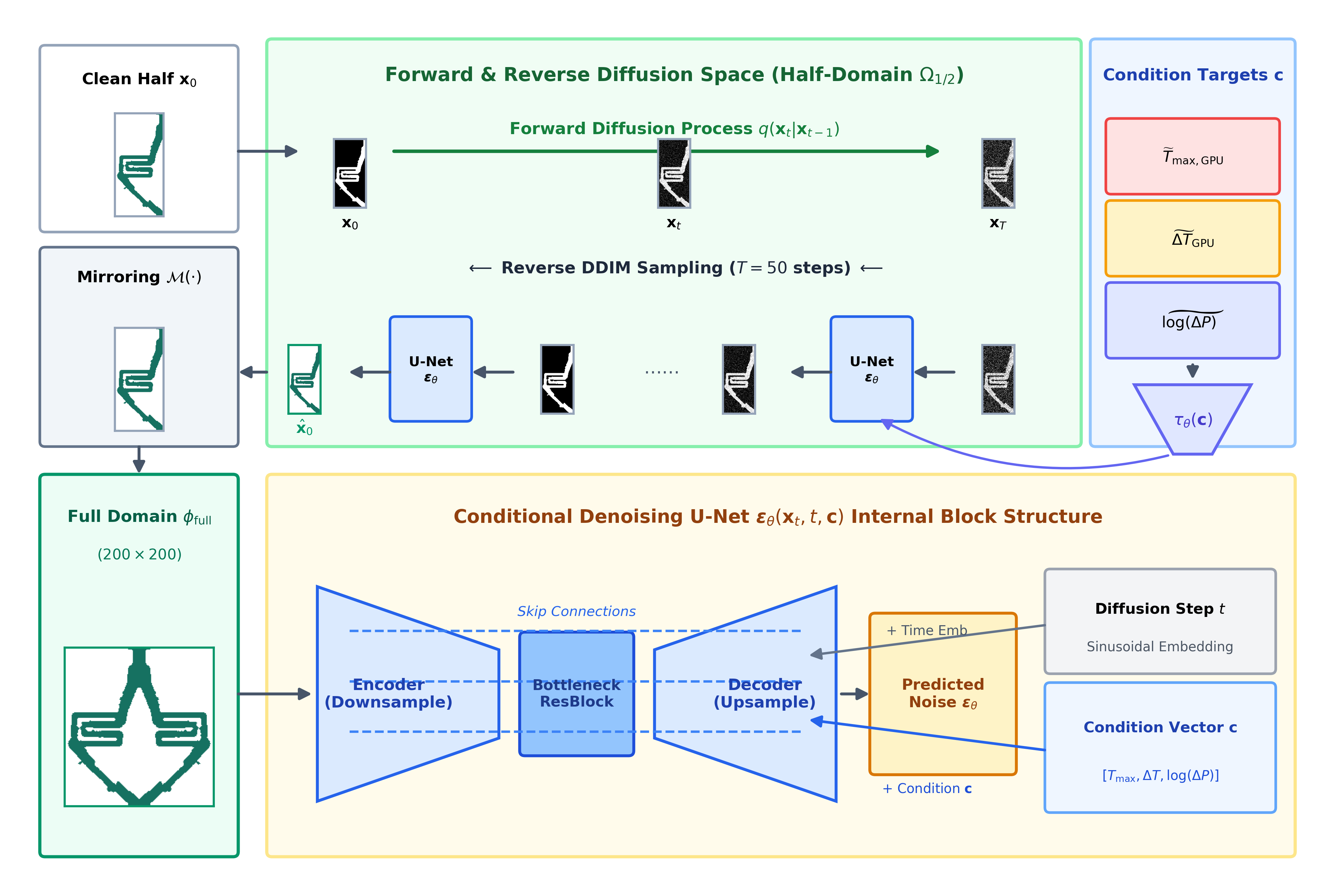}
    \caption{Conditional diffusion framework for symmetric cooling-channel
    topology generation.}
    \label{fig:conditional_diffusion_framework}
\end{figure}

It progressively denoises the sample toward a channel-like topology.
Classifier-free guidance is used during sampling to strengthen the influence of the prescribed performance conditions. The guided noise estimate is expressed in Eq.~\eqref{eq:classifier_free_guidance}.

\begin{equation}
    \hat{\boldsymbol{\epsilon}}_{\theta}
    =
    \boldsymbol{\epsilon}_{\theta}(\mathbf{x}_t,t,\varnothing)
    +
    s_{\mathrm{cfg}}
    \left[
        \boldsymbol{\epsilon}_{\theta}(\mathbf{x}_t,t,\mathbf{c})
        -
        \boldsymbol{\epsilon}_{\theta}(\mathbf{x}_t,t,\varnothing)
    \right]
    \label{eq:classifier_free_guidance}
\end{equation}
where $\varnothing$ denotes the unconditional input and $s_{\mathrm{cfg}}$ is the classifier-free guidance scale. A value of $s_{\mathrm{cfg}}=2.0$ is used in the final generation procedure. The conditional diffusion and symmetry reconstruction framework is summarized in Fig.~\ref{fig:conditional_diffusion_framework}.

To reduce the computational cost of reverse diffusion, deterministic denoising diffusion implicit model (DDIM) sampling is employed~\cite{ho2020denoising, song2020denoising}. The original 200-step diffusion process is sampled using 50 DDIM reverse steps. Deterministic sampling provides reproducible channel generation given a random initialization and a conditioning vector, while substantially reducing the number of denoising evaluations required compared to the full reverse process.

After the reverse diffusion procedure, the continuous network output is converted to a binary half-domain channel mask through thresholding and geometric cleanup. Let the generated half-domain mask be $\phi_{1/2}(x,y)$. The complete channel topology is reconstructed through Eq.~\eqref{eq:symmetry_reconstruction}.

\begin{equation}
    \phi_{\mathrm{full}}
    =
    \left[
        \phi_{1/2}, \,
        \mathcal{M}\left(\phi_{1/2}\right)
    \right]
    \label{eq:symmetry_reconstruction}
\end{equation}
where $\mathcal{M}(\cdot)$ represents reflection about the vertical package centerline. The required inlet and outlet regions are then imposed on the reconstructed mask. Consequently, left-right symmetry is satisfied exactly for every sampled topology.

The conditional diffusion model is intended to efficiently explore the channel-topology design space rather than to replace physics-based feasibility assessment. Although performance conditioning and half-domain generation strongly constrain the generated distribution, they do not guarantee inlet-to-outlet connectivity or the absence of undesirable dead-end branches. All generated designs are therefore passed to the independent topology-screening and thermal-fluids evaluation framework described in Section~\ref{sec:evaluation_framework}.

\section{Design Evaluation and Multi-Objective Selection Framework}
\label{sec:evaluation_framework}


The conditional diffusion model rapidly produces candidate channel topologies, but the generated binary mask is not automatically suitable for physical evaluation. In particular, generative sampling can produce disconnected flow regions or unnecessary dead-end branches that do not contribute to useful inlet-to-outlet coolant transport. A three-stage evaluation framework is therefore applied after generation. First, the sampled masks are screened using topology-based feasibility criteria. Second, all retained designs are evaluated using the calibrated reduced-order thermal-fluids model described in Section~\ref{sec:rom}. Finally, the feasible population is compared using the three design objectives $T_{\max,\mathrm{GPU}}$, $\Delta T_{\mathrm{GPU}}$, and $\Delta P$ to identify non-dominated high-performance candidates for full-order CFD validation.

\subsection{Post-Generation Topology Screening}
\label{subsec:post_generation_screening}

A total of 5{,}000 full-domain channel layouts were sampled from the trained conditional diffusion model. Although symmetry is satisfied by construction through the half-domain generation and reflection procedure described in Section~\ref{subsec:half_domain_generation}, hydraulic connectivity is not guaranteed by the diffusion process. Each generated mask is therefore subjected to a deterministic topology-screening procedure before thermal-fluids evaluation.

The first screening criterion requires a continuous fluid component connecting the prescribed inlet and outlet regions. Connected-component analysis is performed on each binary mask~\cite{rosenfeld1966sequential}, and a topology is retained only if at least one fluid component intersects both ports. Designs without a continuous inlet-to-outlet pathway are rejected because they cannot support the required through-flow.

Of the 5{,}000 generated layouts, 2{,}220 contained a continuous inlet-to-outlet path. The remaining 2{,}780 designs were rejected at this stage. The connected designs were subsequently evaluated for dead-end branching using the branch ratio introduced in Eq.~\eqref{eq:branch_ratio_screening},
\begin{equation}
    F_{\mathrm{branch}}
    =
    \frac{L_{\mathrm{dead}}}{L_{\mathrm{skel}}}
    \label{eq:branch_ratio_screening}
\end{equation}
where $L_{\mathrm{dead}}$ is the total skeleton length belonging to branches that terminate without contributing to the primary through-flow network, and $L_{\mathrm{skel}}$ is the total skeleton length of the channel topology.
The final post-generation feasibility criterion is shown in Eq.~\eqref{eq:branch_constraint}.
\begin{equation}
    F_{\mathrm{branch}} \le F_{\max}, \qquad F_{\max}=0.05
    \label{eq:branch_constraint}
\end{equation}

This criterion removes highly branched layouts that contain substantial stagnant or weakly useful channel regions, while allowing small secondary branches that do not significantly alter the overall through-flow structure. Of the 2,220 connected designs, 1,991 exceeded the branch-ratio threshold and were rejected. The screening procedure therefore retained 229 feasible channel topologies.
It corresponds to a final retention rate of 4.6\% of the generated population. The retained masks were also checked for uniqueness before performance evaluation; all 229 feasible layouts were unique.
No thermal or pressure-drop criterion is applied during this topology-screening stage. Consequently, geometrically feasible designs are not removed simply because their thermal or hydraulic performance is unfavorable. Instead, these quantities are evaluated subsequently and treated explicitly as design objectives.

\subsection{Reduced-Order Evaluation of Feasible Designs}
\label{subsec:rom_evaluation_feasible}

Each of the 229 retained channel masks is evaluated using the same calibrated thermal-fluids ROM described in Section~\ref{sec:rom}. Direct ROM evaluation ensures that all final candidate rankings are based on a single, physically consistent evaluation pathway.
For a channel topology $\phi_i$, the ROM returns a temperature field $T_i(x,y)$ together with the pressure-drop estimate $\Delta P_i$. The thermal performance of each design is evaluated specifically over the two GPU footprints because the GPUs represent the dominant thermal loads in the package. The first objective is the maximum GPU temperature, as shown in Eq.~\eqref{eq:evaluation_tmax}.
\begin{equation}
    T_{\max,\mathrm{GPU},i}
    =
    \max_{(x,y)\in\Omega_{\mathrm{GPU}}} T_i(x,y)
    \label{eq:evaluation_tmax}
\end{equation}
where $\Omega_{\mathrm{GPU}}$ denotes the union of the left and right GPU regions.
The second objective measures temperature nonuniformity across the GPU regions, as shown in Eq.~\eqref{eq:evaluation_deltaT}.
\begin{equation}
    \Delta T_{\mathrm{GPU},i}
    =
    \max_{(x,y)\in\Omega_{\mathrm{GPU}}} T_i(x,y)
    -
    \min_{(x,y)\in\Omega_{\mathrm{GPU}}} T_i(x,y)
    \label{eq:evaluation_deltaT}
\end{equation}

Minimizing $\Delta T_{\mathrm{GPU}}$ discourages designs that reduce the peak temperature at the expense of strong spatial thermal gradients or uneven cooling between the high-power devices.

The third objective is the inlet-to-outlet pressure drop, as shown in Eq.~\eqref{eq:evaluation_dp}.
\begin{equation}
    \Delta P_i = P_{\mathrm{in},i} - P_{\mathrm{out},i}
    \label{eq:evaluation_dp}
\end{equation}

The final design problem can therefore be written in vector form as Eq.~\eqref{eq:three_objective_problem}.
\begin{equation}
    \min_{\phi_i\in\mathcal{F}} \; \mathbf{f}(\phi_i)
    =
    \left[
        T_{\max,\mathrm{GPU},i}, \;
        \Delta T_{\mathrm{GPU},i}, \;
        \Delta P_i
    \right]
    \label{eq:three_objective_problem}
\end{equation}
where $\mathcal{F}$ denotes the set of topologically feasible generated designs satisfying the connectivity and branch-ratio constraints.
All three objectives are minimized. The first objective suppresses the highest GPU junction temperature, the second promotes temperature uniformity, and the third limits hydraulic resistance. These objectives are retained independently rather than immediately combined into a single weighted objective, so that the thermal-fluids trade-offs present in the generated design population remain visible.

\subsection{Multi-Objective Analysis and Candidate Selection}
\label{subsec:multiobjective_selection}

The ROM-evaluated candidate population is analyzed using Pareto dominance. For two feasible designs $\phi_a$ and $\phi_b$, design $\phi_a$ is defined to dominate $\phi_b$ if
\begin{equation}
    f_j(\phi_a) \le f_j(\phi_b) \qquad \forall j\in\{1,2,3\}
    \label{eq:pareto_weak}
\end{equation}
and
\begin{equation}
    f_k(\phi_a) < f_k(\phi_b) \qquad \text{for at least one } k
    \label{eq:pareto_strict}
\end{equation}

A topology is therefore Pareto optimal if no other feasible generated design improves at least one of the three objectives without worsening another. The resulting non-dominated set represents the performance boundary of the generated design population and exposes the trade-offs between hotspot suppression, temperature uniformity, and hydraulic resistance.

For visualization, the principal Pareto projection is presented in the $T_{\max,\mathrm{GPU}}$--$\Delta P$ plane, while $\Delta T_{\mathrm{GPU}}$ is represented using the marker color. This 2D projection provides a clear representation of the dominant thermal-fluids trade-off while retaining information from the third objective. The underlying candidate assessment nevertheless considers all three metrics.

Because the final application places particular emphasis on preventing excessive GPU junction temperature, the final candidate is selected from the feasible non-dominated population with priority given to the lowest $T_{\max,\mathrm{GPU}}$, while also requiring low $\Delta T_{\mathrm{GPU}}$, acceptable hydraulic performance, and a physically clean connected topology. This decision rule avoids selecting a design solely through an arbitrary weighted sum of the three quantities.

\section{Results and Discussion}
\label{sec:results}

\subsection{Generated-Design Feasibility and Topology Screening}
\label{subsec:generation_results}

The trained conditional diffusion model was used to generate 5,000 new symmetric cooling-channel layouts. The generated population exhibited substantial geometric diversity, including differences in channel spreading, branching, flow-path width, and the routing of the coolant across the GPU and CPU regions. Because left-right symmetry was imposed through half-domain generation and mirroring, all sampled layouts satisfied the prescribed symmetry requirement by construction.

The principal limitation of the raw generated population was hydraulic connectivity. Of the 5,000 generated layouts, 2,220 contained a continuous coolant pathway connecting the prescribed inlet and outlet, corresponding to a connectivity rate of $44.4\%$. The remaining 2,780 layouts were rejected because no complete inlet-to-outlet flow path was present.

\begin{figure}[h]
    \centering
    \includegraphics[width=\linewidth]{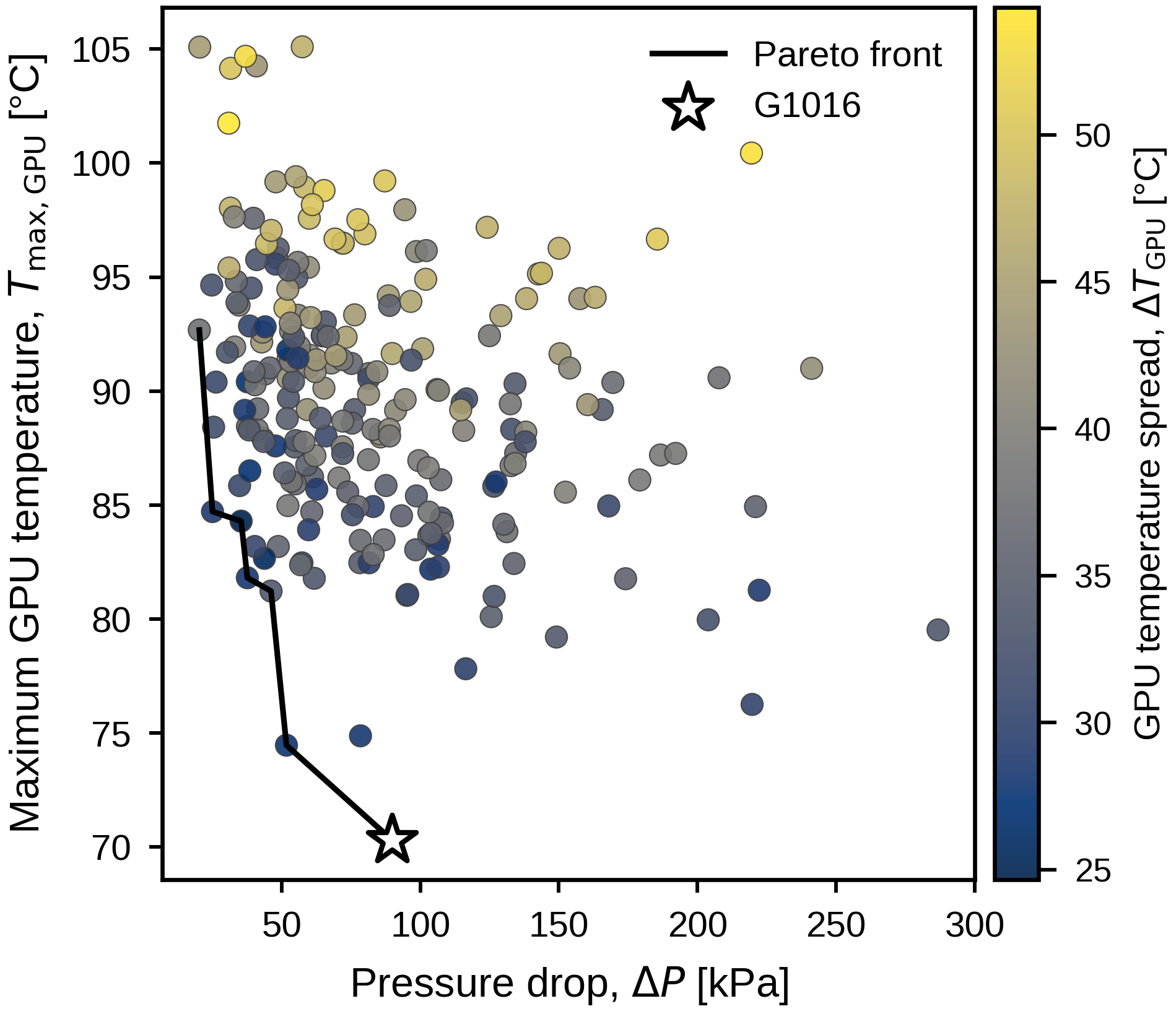}
    \caption{Thermal-fluids performance of the feasible generated
    channel layouts.}
    \label{fig:pareto_population}
\end{figure}

The connected layouts were further screened using the dead-end branch-ratio criterion, $F_{\mathrm{branch}}\leq0.05$. Of the 2{,}220 connected designs, 1{,}991 exceeded the allowable branch ratio and were removed. A total of 229 designs therefore satisfied the complete post-generation topology-screening procedure, corresponding to $4.58\%$ of the original generated population. All 229 retained masks were geometrically unique.

These results indicate that the diffusion model can generate a broad range of channel-like geometries, but generative sampling alone is insufficient to ensure hydraulic feasibility. In particular, connectivity and excessive branching remain important failure modes. The deterministic physics-guided screening stage is therefore essential to prevent nonfunctional layouts from entering the thermal-fluids evaluation stage.

\begin{figure}[t]
    \centering
    \includegraphics[
        width=0.82\textwidth
    ]{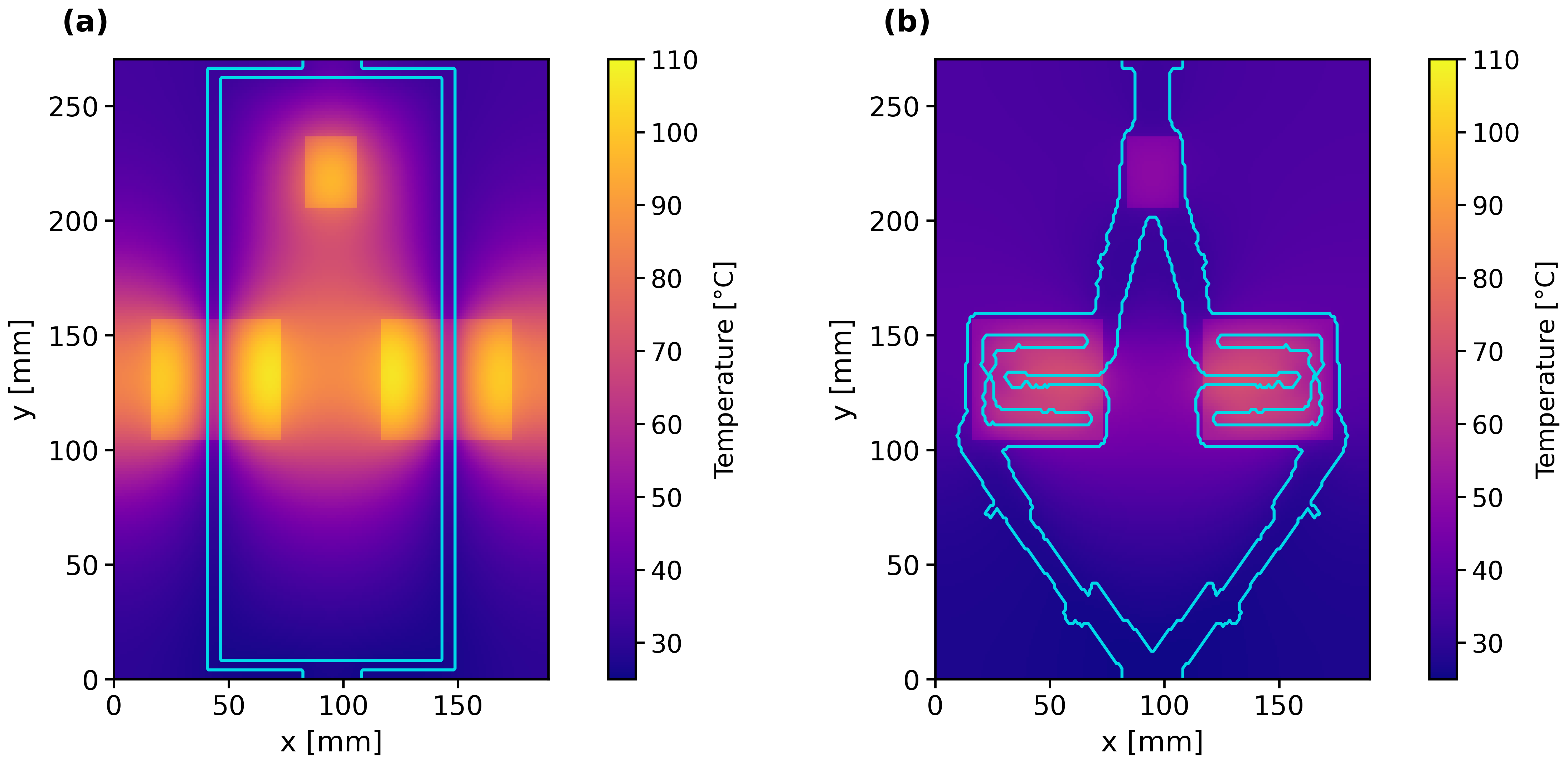}
    \caption{ROM-estimated temperature distributions for
    (a) the conventional reference topology and
    (b) the optimal topology under identical operating
    conditions.}
    \label{fig:s1_g1016_comparison}
\end{figure}

\subsection{Thermal-Fluids Performance of the Feasible Design Population}
\label{subsec:population_results}

The 229 topologically feasible channel designs were subsequently evaluated using the calibrated reduced-order thermal-fluids model. Considerable variation was observed across all three design objectives, demonstrating that topological changes strongly influence both thermal performance and hydraulic resistance.
Across the feasible population, the estimated maximum GPU temperature ranged from $70.30^\circ\mathrm{C}$ to $105.10^\circ\mathrm{C}$, with a mean value of $89.64^\circ\mathrm{C}$ and a standard deviation of $5.72^\circ\mathrm{C}$. The GPU temperature spread ranged from $24.66^\circ\mathrm{C}$ to $54.34^\circ\mathrm{C}$, with a mean of $38.05^\circ\mathrm{C}$ and a standard deviation of $6.16^\circ\mathrm{C}$. The mean estimated pressure drop across the feasible population was $84.17~\mathrm{kPa}$.

Figure~\ref{fig:pareto_population} presents the feasible design population in the $T_{\max,\mathrm{GPU}}$--$\Delta P$ objective plane, with $\Delta T_{\mathrm{GPU}}$ represented by marker color. The population does not collapse onto a single optimum; instead, different topologies occupy different regions of the thermal-fluids performance space. Designs with lower hydraulic resistance do not necessarily provide the lowest peak GPU temperature, while designs with strong hotspot suppression may require more restrictive coolant pathways.


The Pareto front identifies channel layouts for which improving one of the displayed objectives cannot be achieved without deteriorating another. This behavior confirms that the generated design problem remains fundamentally multi-objective, even though some individual generated layouts substantially outperform conventional configurations.
Among the 229 feasible layouts, optimal design (G1016) achieved the lowest estimated maximum GPU temperature, $T_{\max,\mathrm{GPU}} = 70.30^\circ\mathrm{C}$. The corresponding GPU temperature spread was $\Delta T_{\mathrm{GPU}} = 24.90^\circ\mathrm{C}$, and the estimated pressure drop was $\Delta P = 89.72~\mathrm{kPa}$.

Although the optimal design does not individually minimize all three objectives, it lies on the $T_{\max,\mathrm{GPU}}$--$\Delta P$ Pareto front and provides the strongest peak-temperature performance of the feasible population while retaining low temperature nonuniformity and moderate hydraulic resistance. Its branch ratio, $F_{\mathrm{branch}}=0.0306$, also remains below the prescribed post-generation limit of $0.05$. The optimal design was therefore selected as the final topology for comparison against the conventional reference design and for independent full-order CFD validation.

\subsection{Comparison With the Conventional Reference Design}
\label{subsec:baseline_comparison}

The selected generated topology was compared with the conventional straight-channel reference design, denoted S1, under identical thermal loads, coolant conditions, material properties, and package dimensions. Figure~\ref{fig:s1_g1016_comparison} compares the ROM-estimated temperature distributions of the two designs.

The conventional S1 topology produced a maximum GPU temperature of $105.89^\circ\mathrm{C}$ and a GPU temperature spread of $52.40^\circ\mathrm{C}$. In comparison, the optimal design reduced the maximum GPU temperature to $70.30^\circ\mathrm{C}$, and the temperature spread to $24.90^\circ\mathrm{C}$, corresponding to reductions of $33.6\%$ and $52.5\%$, respectively. The generated topology also provided a substantial hydraulic improvement. The estimated pressure drop of the S1 reference design was $329.40~\mathrm{kPa}$, whereas the optimal design required only $89.72~\mathrm{kPa}$, corresponding to a pressure-drop reduction of $72.8\%$. The comparison is summarized in Table~\ref{tab:s1_g1016}.

\begin{table}[htbp]
\centering
\caption{Thermal-fluids comparison between the conventional S1 reference topology and the optimal design.}
\label{tab:s1_g1016}

\begin{tabular}{lccc}
\hline
Metric & S1 & optimal design & Reduction \\
\hline
$T_{\max,\mathrm{GPU}}$ [$^\circ$C] & 105.89 & 70.30 & 33.6\% \\
$\Delta T_{\mathrm{GPU}}$ [$^\circ$C] & 52.40 & 24.90 & 52.5\% \\
$\Delta P$ [kPa] & 329.40 & 89.72 & 72.8\% \\
\hline
\end{tabular}
\end{table}

The improvement in optimal design is not produced by a simple increase in coolant-channel area. Instead, the generated topology redistributes the available flow paths so that coolant is directed more effectively across the high-power GPU regions while avoiding the long, restrictive flow path characteristic of the conventional reference topology. The resulting geometry simultaneously improves heat removal and reduces hydraulic resistance.

\begin{figure}[t]
    \centering
    \includegraphics[
        width=0.95\textwidth
    ]{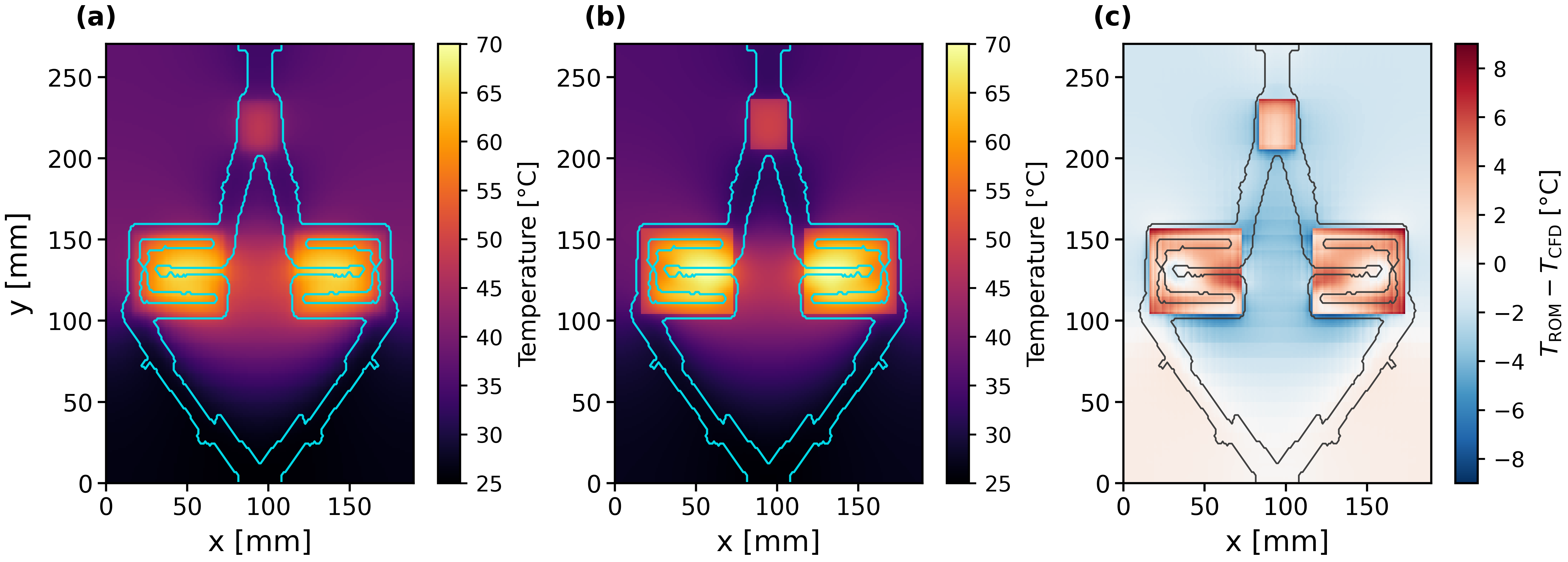}
    \caption{Comparison of the optimal topology using
    (a) the full-order CFD temperature field,
    (b) the reduced-order model prediction, and
    (c) the temperature difference
    $T_{\mathrm{ROM}}-T_{\mathrm{CFD}}$.
    }
    \label{fig:g1016_rom_cfd_validation}
\end{figure}

This result is particularly significant because thermal and hydraulic objectives are generally competing: increasing local coolant access can introduce additional flow resistance, while reducing resistance can encourage coolant bypass around the heat-generating regions. The optimal design demonstrates that topology modification can instead improve both effects simultaneously relative to the conventional reference design. The broader generated population nevertheless retains the expected thermal-fluids trade-offs shown in Fig.~\ref{fig:pareto_population}; therefore, the simultaneous improvement observed for the optimal design should be interpreted relative to the S1 baseline rather than as evidence that the underlying multi-objective trade-off has been eliminated.

\subsection{Full-Order CFD Validation of the Selected Design}
\label{sec:selected_design_validation}

The optimal design topology was reconstructed in 3D and independently evaluated using the full-order OpenFOAM conjugate heat-transfer model described in Section~\ref{sec:cfd_governing}. The ROM calibration coefficients were held fixed during this comparison; therefore, no design-specific recalibration was performed using the optimal design CFD solution.

The CFD solution reached a stable thermal and hydraulic state and satisfied the prescribed mass- and energy-conservation checks. The integrated heat removal through the coolant interface was $2{,}699.2~\mathrm{W}$ for an applied chip power of $2{,}700~\mathrm{W}$, corresponding to an energy imbalance below $0.1\%$. The mass-flow-weighted coolant outlet temperature was $32.79^\circ\mathrm{C}$.

The full order simulation estimated a maximum GPU temperature of $T_{\max,\mathrm{GPU}}^{\mathrm{CFD}} = 66.70^\circ\mathrm{C}$, compared with the ROM prediction of $T_{\max,\mathrm{GPU}}^{\mathrm{ROM}} = 70.30^\circ\mathrm{C}$.
Because both calculations use an inlet coolant temperature of $25^\circ\mathrm{C}$, the relative error is evaluated using temperature rise above the inlet. The ROM and CFD temperature rises are therefore $45.30^\circ\mathrm{C}$ and $41.70^\circ\mathrm{C}$, respectively, giving 8.6\%

The CFD solution produced a GPU temperature spread of $29.7^\circ\mathrm{C}$, compared with $24.90^\circ\mathrm{C}$ estimated by the ROM. The corresponding signed relative difference is 16.1\%.
The temperature-spread prediction therefore exhibits a larger discrepancy than the peak-temperature prediction. This is consistent with the greater sensitivity of $\Delta T_{\mathrm{GPU}}$ to local flow redistribution and 3D heat spreading, which are represented in simplified form by the reduced-order formulation.
The CFD pressure drop was statistically stable at 92.1{kPa} compared with the ROM prediction of $89.72~\mathrm{kPa}$. The corresponding signed pressure-drop error is 2.6\%.

Figure~\ref{fig:g1016_rom_cfd_validation} compares the full-order CFD temperature field with the corresponding reduced-order prediction and the local signed temperature difference. Both models reproduce the same dominant thermal structure, including the principal hot regions over the two GPU footprints and the substantially cooler CPU region. The ROM generally predicts a slightly higher peak temperature, while the largest local discrepancies occur within and downstream of the GPU cooling regions.

Overall, the independent optimal design comparison shows that the calibrated ROM reproduces the two quantities most important for initial topology screening, peak GPU temperature and pressure drop, with errors of $8.6\%$ and $2.6\%$, respectively. The larger error in GPU temperature spread indicates that the ROM is less accurate in resolving detailed spatial temperature nonuniformity. Nevertheless, the agreement is sufficient for its intended role as a rapid comparative screening model, particularly because full 3D CFD is applied only after the design space has been reduced to a small number of promising candidates. The ROM and CFD results are summarized in Table~\ref{tab:g1016_validation}.

\begin{table}[h]
\centering
\caption{Comparison of the reduced-order and full-order predictions for optimal design.}
\label{tab:g1016_validation}

\begin{tabular}{lccc}
\hline
Metric & ROM & CFD & Relative difference \\
\hline
$T_{\max,\mathrm{GPU}}$ [$^\circ$C] & 70.30 & 66.70 & $+8.6\%^{a}$ \\
$\Delta T_{\mathrm{GPU}}$ [$^\circ$C] & 24.90 & $29.7$ & $-16.1\%$ \\
$\Delta P$ [kPa] & 89.72 & $92.1$ & $-2.6\%$ \\
\hline
\multicolumn{4}{l}{$^{a}$Relative error evaluated using temperature rise above $T_{\mathrm{in}}=25^\circ$C.}
\end{tabular}
\end{table}

\subsection{Limitations}
\label{subsec:limitations}

The final feasibility rate of $4.58\%$ shows a primary limitation of the present generative approach. Although symmetry is imposed exactly through half-domain generation, inlet-to-outlet connectivity and low dead-end branching are not guaranteed directly by the diffusion process, therefore, the framework relies on deterministic post-generation filtering. Incorporating connectivity constraints directly into the reverse-sampling procedure or using physics-informed guidance could substantially increase the feasible yield.
The reduced-order representation employs depth-averaged flow, correlation-based convective heat transfer, path-dependent coolant warming, and equivalent in-plane solid conduction. While these approximations enable rapid population-scale evaluation, they cannot capture complex 3D flow phenomena, particularly local recirculation and through-thickness heat spreading. The $-16.1\%$ discrepancy in GPU temperature spread for optimal design is consistent with these simplifications, whereas peak temperature rise and pressure drop, which depend less on resolving detailed local gradients, agree to within $+8.6\%$ and $-2.6\%$, respectively.
The ROM calibration coefficients are held fixed across the entire generated population. While this avoids topology-specific overfitting, it retains dependence on the geometric scales and operating envelope used during calibration. Applying the framework to substantially different flow rates, channel depths, coolant fluids, or package layouts would require recalibration and validation.
Also, the generative model operates on a 2D planform with a uniform extrusion depth. Consequently, the study optimizes in-plane topology rather than local channel depth, cross-sectional profile, or 3D manifold distribution.

\section{Conclusions and Future Work}
\label{sec:conclusion}
This work presents a generative design framework for liquid cooling channel design in a $2.7~\mathrm{kW}$ multi-chip package. The framework combines conditional diffusion-based generation, topology screening, reduced-order thermal-fluids evaluation, multi-objective analysis, and final OpenFOAM validation. 
The optimal design was obtained from 5{,}000 generated layouts, achieving $T_{\max,\mathrm{GPU}}=70.30^\circ\mathrm{C}$, $\Delta T_{\mathrm{GPU}}=24.90^\circ\mathrm{C}$, and $\Delta P=89.72~\mathrm{kPa}$. Compared with the conventional reference, the optimal design reduced the maximum GPU temperature by $33.6\%$, the GPU temperature spread by $52.5\%$, and the pressure drop by $72.8\%$. High-fidelity OpenFOAM validates the optimal design's maximum GPU temperature of $66.70^\circ\mathrm{C}$ and a pressure drop of $92.1~\mathrm{kPa}$, showing good agreement with the reduced-order predictions for the primary thermal and hydraulic metrics.
Overall, the results show that generative design can identify non-conventional cooling-channel layouts that outperform conventional designs while reducing the need for CFD-based trial and error. Future work will focus on improving topology feasibility during generation, extending the design space to 3D channel geometries, and experimentally validating selected designs.

\section*{Acknowledgments}
This research was partially supported by the University of Michigan-Dearborn Office of Research through the Research Initiation \& Development (RID) Grant.

\section*{Data and Code Availability}
The data and code will be made publicly available once the paper is accepted.



\begin{nomenclature}

\EntryHeading{Roman Symbols}

\entry{$A$}{flow cross-sectional area (m$^2$)}
\entry{$c_p$}{specific heat capacity (J kg$^{-1}$ K$^{-1}$)}
\entry{$C_h$}{calibrated heat-transfer correction coefficient}
\entry{$C_p$}{calibrated pressure-drop correction coefficient}
\entry{$d_w$}{Euclidean distance to nearest solid wall (m)}
\entry{$D_h$}{hydraulic diameter (m)}
\entry{$f$}{Darcy friction factor}
\entry{$F_{\mathrm{branch}}$}{dead-end branch ratio}
\entry{$F_{\max}$}{maximum allowable dead-end branch ratio}
\entry{$h$}{local convective heat-transfer coefficient (W m$^{-2}$ K$^{-1}$)}
\entry{$H$}{coolant-channel depth (m)}
\entry{$k$}{thermal conductivity (W m$^{-1}$ K$^{-1}$)}
\entry{$k_{\mathrm{eff}}$}{effective in-plane thermal conductivity (W m$^{-1}$ K$^{-1}$)}
\entry{$K$}{minor-loss coefficient}
\entry{$L_{\mathrm{dead}}$}{total dead-end branch skeleton length (m)}
\entry{$L_{\mathrm{path}}$}{connected coolant-path length (m)}
\entry{$L_{\mathrm{skel}}$}{total channel-skeleton length (m)}
\entry{$L_x, L_y$}{computational domain dimensions (m)}
\entry{$\dot{m}$}{coolant mass-flow rate (kg s$^{-1}$)}
\entry{$N_x, N_y$}{grid resolution along coordinate axes}
\entry{$p$}{static pressure (Pa)}
\entry{$\Delta P$}{inlet-to-outlet pressure drop (Pa)}
\entry{$q''$}{surface heat flux (W m$^{-2}$)}
\entry{$q'''$}{volumetric heat-generation rate (W m$^{-3}$)}
\entry{$\dot{Q}$}{heat-transfer rate (W)}
\entry{$R_{\mathrm{stack}}$}{through-thickness thermal resistance (m$^2$ K W$^{-1}$)}
\entry{$s$}{normalized coolant-path progress variable}
\entry{$s_{\mathrm{cfg}}$}{classifier-free guidance scale}
\entry{$t_c$}{coolant-channel depth (m)}
\entry{$t_{\mathrm{eff}}$}{effective solid-stack thickness (m)}
\entry{$t_{\mathrm{Cu}}, t_{\mathrm{Si}}$}{layer thickness of copper and silicon (m)}
\entry{$T$}{temperature (K or $^\circ$C)}
\entry{$\Delta T_{\mathrm{GPU}}$}{GPU temperature spread ($^\circ$C)}
\entry{$T_{\max,\mathrm{GPU}}$}{maximum GPU temperature ($^\circ$C)}
\entry{$T_{\min,\mathrm{GPU}}$}{minimum GPU temperature ($^\circ$C)}
\entry{$\mathbf{u}$}{coolant velocity vector (m s$^{-1}$)}
\entry{$V$}{mean coolant velocity (m s$^{-1}$)}
\entry{$\dot{V}$}{volumetric flow rate (m$^3$ s$^{-1}$ or L min$^{-1}$)}
\entry{$w$}{local channel width (m)}
\entry{$w_i$}{topology-dependent diffusion-training weight}
\entry{$\mathbf{x}_0$}{clean binary channel-mask sample}
\entry{$\mathbf{x}_t$}{noised diffusion sample at time step $t$}

\EntryHeading{Greek Letters}

\entry{$\alpha_t$}{diffusion retention coefficient at time step $t$}
\entry{$\bar{\alpha}_t$}{cumulative diffusion retention coefficient}
\entry{$\beta_t$}{diffusion variance schedule value at time step $t$}
\entry{$\gamma$}{flow-regime transition weighting factor}
\entry{$\boldsymbol{\epsilon}$}{standard Gaussian noise vector}
\entry{$\boldsymbol{\epsilon}_{\theta}$}{noise vector estimated by denoising network}
\entry{$\varepsilon$}{relative percentage error (\%)}
\entry{$\mu$}{dynamic viscosity (Pa s)}
\entry{$\mu_t$}{turbulent eddy viscosity (Pa s)}
\entry{$\rho$}{density (kg m$^{-3}$)}
\entry{$\tau$}{channel tortuosity}
\entry{$\phi$}{binary channel planform mask indicator}
\entry{$\theta$}{trainable diffusion-model parameter set}
\entry{$\Omega$}{spatial domain}
\entry{$\omega$}{specific turbulence dissipation rate (s$^{-1}$)}

\EntryHeading{Dimensionless Groups}

\entry{$Nu$}{Nusselt number, $h D_h / k_f$}
\entry{$Pr$}{Prandtl number, $\mu c_p / k_f$}
\entry{$Re$}{Reynolds number, $\rho V D_h / \mu$}

\EntryHeading{Superscripts and Subscripts}

\entry{$\mathrm{abs}$}{absorbed by the coolant}
\entry{$\mathrm{amb}$}{ambient condition}
\entry{$\mathrm{CFD}$}{full-order conjugate heat-transfer prediction}
\entry{$\mathrm{con}$}{contraction minor loss}
\entry{$\mathrm{CPU}$}{central processing unit}
\entry{$\mathrm{eff}$}{effective property}
\entry{$\mathrm{exp}$}{expansion minor loss}
\entry{$\mathrm{f}$}{fluid or coolant domain}
\entry{$\mathrm{fric}$}{frictional contribution}
\entry{$\mathrm{GPU}$}{graphics processing unit}
\entry{$\mathrm{in}$}{inlet condition}
\entry{$\mathrm{lam}$}{laminar flow regime}
\entry{$\max$}{maximum value}
\entry{$\min$}{minimum value}
\entry{$\mathrm{minor}$}{minor-loss contribution}
\entry{$\mathrm{mw}$}{mass-flow-weighted average}
\entry{$\mathrm{out}$}{outlet condition}
\entry{$\mathrm{p}$}{effective plate condition}
\entry{$\mathrm{ROM}$}{reduced-order-model prediction}
\entry{$\mathrm{s}$}{solid domain}
\entry{$\mathrm{t}$}{turbulent flow regime}
\entry{$\mathrm{turb}$}{turbulent correlation}

\EntryHeading{Abbreviations}

\entry{CFD}{computational fluid dynamics}
\entry{CHT}{conjugate heat transfer}
\entry{CPU}{central processing unit}
\entry{DDIM}{denoising diffusion implicit model}
\entry{GPU}{graphics processing unit}
\entry{MCM}{multi-chip module}
\entry{RANS}{Reynolds-averaged Navier--Stokes}
\entry{ROM}{reduced-order model}
\entry{SST}{shear-stress transport}

\end{nomenclature}







\end{document}